\documentclass[a4paper,12pt]{article}
\pdfoutput=1
\usepackage{graphicx,rotating,colortbl}
\usepackage{hyperref,slashed,amsfonts}
\usepackage{color}
\usepackage{amsmath}
\hypersetup{colorlinks,bookmarksopen,bookmarksnumbered,
linkcolor=blus,pdfstartview=FitH,urlcolor=rossos,citecolor=verde}

\newcommand{\NF}{N_{\rm TF}}
\newcommand{\Z}{\mathbb{Z}}

\usepackage{lscape}
\usepackage{tikzfeynman}
\usepackage{youngtab}
\usepackage{multicol}
\usepackage{color}
\definecolor{rosso}{cmyk}{0,1,1,0.4}
\definecolor{rossos}{cmyk}{0,1,1,0.55}
\definecolor{rossoc}{cmyk}{0,1,1,0.2}
\definecolor{blu}{cmyk}{1,1,0,0.3}
\definecolor{blus}{cmyk}{1,1,0,0.6}
\definecolor{bluc}{cmyk}{1,1,0,0.1}
\definecolor{verde}{cmyk}{0.92,0,0.59,0.25}
\definecolor{verdec}{cmyk}{0.92,0,0.59,0.15}
\definecolor{verdes}{cmyk}{0.92,0,0.59,0.4}

\newcommand{\Q}{\Psi}

\newcommand{\eq}[1]{~{\rm (\ref{eq:#1})}}

\newcommand{\MeV}{\,{\rm MeV}}
\newcommand{\GeV}{\,{\rm GeV}}
\newcommand{\TeV}{\,{\rm TeV}}
\newcommand{\cm}{\,{\rm cm}}
\newcommand{\Tr}{\,{\rm Tr}}

\def\circa#1{\,\raise.3ex\hbox{$#1$\kern-.75em\lower1ex\hbox{$\sim$}}\,}

\newcommand{\beq}{\begin{equation}}
\newcommand{\eeq}{\end{equation}}

\newcommand{\bea}{\begin{eqnarray}}
\newcommand{\eea}{\end{eqnarray}}
\newcommand{\be}{\begin{equation}}
\newcommand{\ee}{\end{equation}}
\font\tenrsfs=rsfs10 at 12pt
\font\sevenrsfs=rsfs7
\font\fiversfs=rsfs5
\newfam\rsfsfam
\textfont\rsfsfam=\tenrsfs
\scriptfont\rsfsfam=\sevenrsfs
\scriptscriptfont\rsfsfam=\fiversfs
\def\mathscr#1{{\fam\rsfsfam\relax#1}}

\def\Lag{\mathscr{L}}

\def\circa#1{\,\raise.3ex\hbox{$#1$\kern-.75em\lower1ex\hbox{$\sim$}}\,}
\makeatletter

\def\hhref#1{\href{http://arxiv.org/abs/#1}{arXiv:#1}} 
\def\arXiv#1{\href{http://arxiv.org/abs/#1}{arXiv:#1}} 

\def\art{\@ifnextchar[{\eart}{\oart}}
\def\eart[#1]#2#3#4#5#6{{\rm #2}, {\em #3 \bf #4} {\rm (#6) #5} ({\em #1})}
\def\hepart[#1]#2{{\rm #2, \hhref{#1}}}
\newcommand{\oart}[5]{{\rm #1}, {\em #2 \bf #3} {\rm (#5) #4}}

\newcounter{alphaequation}[equation]
\def\thealphaequation{\theequation\hbox to
0.6em{\hfil\alph{alphaequation}\hfil}}
\def\eqnsystem#1{
\def\@eqnnum{{\rm (\thealphaequation)}}
\def\@@eqncr{\let\@tempa\relax \ifcase\@eqcnt \def\@tempa{& & &} \or
  \def\@tempa{& &}\or \def\@tempa{&}\fi\@tempa
  \if@eqnsw\@eqnnum\refstepcounter{alphaequation}\fi
\global\@eqnswtrue\global\@eqcnt=0\cr}
\refstepcounter{equation} \let\@currentlabel\theequation \def\@tempb{#1}
\ifx\@tempb\empty\else\label{#1}\fi
\refstepcounter{alphaequation}
\let\@currentlabel\thealphaequation
\global\@eqnswtrue\global\@eqcnt=0 \tabskip\@centering\let\\=\@eqncr
$$\halign to \displaywidth\bgroup \@eqnsel\hskip\@centering
$\displaystyle\tabskip\z@{##}$&\global\@eqcnt\@ne
\hskip2\arraycolsep\hfil${##}$\hfil& \global\@eqcnt\tw@\hskip2\arraycolsep
$\displaystyle\tabskip\z@{##}$\hfil
\tabskip\@centering&\llap{##}\tabskip\z@\cr}
\def\endeqnsystem{\@@eqncr\egroup$$\global\@ignoretrue} \makeatother

\newcommand{\SO}{\,{\rm SO}}
\newcommand{\SU}{\,{\rm SU}}
\newcommand{\U}{\,{\rm U}}
\begin{document}
\centerline{IFUP-TH/2015}

\vspace{2cm}

\begin{center}
{\Huge \bf \color{rossos}
Accidental Composite Dark Matter}\\[1cm]

{\bf Oleg Antipin$^{a}$, Michele Redi$^{a}$, Alessandro Strumia$^{b,c}$, Elena Vigiani$^{b}$}  
\\[7mm]

{\it $^a$INFN, Sezione di Firenze, Via G. Sansone, 1; I-50019 Sesto Fiorentino, Italy}\\
{\it $^b$ Dipartimento di Fisica dell'Universit{\`a} di Pisa and INFN, Italy}\\
{\it $^c$ National Institute of Chemical Physics and Biophysics, Tallinn, Estonia}

\vspace{1cm}
{\large\bf\color{blus} Abstract}
\begin{quote}\large
We build models where Dark Matter candidates arise as composite states of a new confining gauge force,
stable thanks to accidental symmetries. Restricting to renormalizable theories compatible with SU(5) unification, we find 13 
models based on SU($N$) gauge theories and 9  based on SO($N$).
We also describe other models that require non-renormalizable interactions.
The two gauge groups lead to distinctive phenomenologies: $\SU(N)$  theories give complex DM, with potentially observable 
electric and magnetic dipole moments that lead to peculiar spin-independent cross sections;
$\SO(N)$  theories give real DM, with challenging spin-dependent cross sections or inelastic scatterings. 
Models with Yukawa couplings also give rise to
spin-independent direct detection mediated by the Higgs boson and to electric dipole moments for the electron.
In some models DM has higher spin. Each model predicts a specific set of lighter composite scalars, possibly observable at colliders.

\end{quote}

\thispagestyle{empty}
\end{center}
\begin{quote}
{\large\noindent\color{blus} 
}

\end{quote}
\vspace{-1.5cm}

\newpage

\tableofcontents

\setcounter{footnote}{0}


\section{Introduction}

A striking success of the Standard Model is that all observed global symmetries are understood as accidental symmetries of the renormalizable
Lagrangian. This explains in particular the stability of the proton as a consequence of  baryon number conservation. 

In nature, besides the proton, at least another particle should be stable to provide the necessary Dark Matter (DM) abundance required by cosmological 
observations.  It is natural to imagine that  dark matter too is stable because of  accidental symmetries. 
This idea can be minimally realized by adding to the SM one extra multiplet  that cannot have any Yukawa interaction with SM particles,
and that contains a DM candidate~\cite{MDM}.

The fact that bounds from DM searches require a successful weak-scale
DM candidate to have no electric charge, no color, and almost no coupling to the $Z$
(the vectorial coupling to the $Z$ must be a few orders of magnitude smaller than a typical weak coupling) calls for an explanation.
A simple way of explaining why DM is so dark and stable is to add to the SM (with its elementary Higgs) 
new fermions $\Q$ charged under a new technicolor interaction that confines at a scale $\Lambda_{\rm TC}$.
Techni-quarks are assumed to lie in (possibly reducible) real representations under the SM gauge group, such that their condensates do not break 
the electro-weak symmetry, realising the framework dubbed `vector-like confinement' in \cite{sundrum}. 
The renormalizable Lagrangian of the theory is
\beq 
\Lag = \Lag_{\rm SM}    +  \bar\Q_i( i\slashed{D}  - m_i ) \Q_i - \frac{{{\cal G}_{\mu\nu}^{A2} }}{4g_{\rm TC}^2} 
+\frac {\theta_{\rm TC}} {32\pi^2} {\cal G}^A_{\mu\nu}\tilde{\cal G}^A_{\mu\nu}
+ [ H\bar\Q_i(y^L_{ij}P_L+ y^R_{ij}P_R) \Q_j + \hbox{h.c.}]
\label{lagrangian}
\eeq
where the latter term,  Yukawa interactions with the Higgs doublet $H$, can be allowed by quantum numbers.
The  topological term for  technicolor gauge fields is physical for non-vanishing techni-quark masses $m_i$.

We assume that when technicolor interactions confine at a scale $\Lambda_{\rm TC}$, the approximate global techni-flavor symmetry
is broken by condensates producing light techni-pions (TC$\pi$) and other heavier composite particles, such as techni-baryons (TCb).
All these particles are splitted in mass by SM gauge interactions in such a way that the lightest stable techni-particle 
(charged under an accidental symmetry that keeps it stable) tends to be the  `most neutral' one.

\smallskip

Composite Dark Matter has been rarely considered in the literature,  and mostly 
in models  with different goals, e.g.\ with supersymmetry~\cite{CDM-SUSY}, with composite~\cite{CDM-TC} or partially composite Higgs~\cite{CDM-PC},
with a mirror-SM sector~\cite{Khlopov} or quirks~\cite{quirks} or a fourth generation~\cite{CDM-4g}
as well as from a phenomenological point of view,  in order to realise special situations
(such as inelastic DM, asymmetric DM, strongly interacting DM, magnetic DM, etc.) often motivated by anomalies~\cite{CDM-sigma}.
An approach  similar to the present study  was considered in \cite{Appelquist:2014jch,us,Appelquist}. 
In~\cite{Appelquist:2014jch,Appelquist} bosonic techni-baryon DM in $\SU(4)$ gauge theories was studied. 
In~\cite{us} we began a general study of composite DM adopting a specific point of view with respect
to the naturalness problem, according to which the Lagrangian does not contain any massive parameter,
power divergences are unphysical,  all masses arise via dimensional transmutation.
The resulting assumption \mbox{$m_i=0$}  lead to  very predictive models~\cite{us}.
Allowing  for techni-quark masses (if lighter than about 1 TeV, they do not induce unnaturally large corrections to the Higgs mass~\cite{FinNat})
and for an order one $\theta_{\rm TC}$ modifies the mass spectrum of the theory, inducing 
electric dipole moments (EDMs) for TCb that leads to a sizeable Dark Matter direct detection signal
with characteristic dependence on velocity and transferred momentum.

\medskip

The issue of composite dark matter is logically independent from the point of view in \cite{us} on naturalness.
We here revisit the DM issue remaining agnostic about the explanation of smallness of the electro-weak scale: 
we just assume that for some reason the SM is much lighter than other unspecified new physics, such that accidental symmetries appear at low energy.
We make the following simplifying  assumptions:
\begin{enumerate}
\item  We study both $\SU(N)_{\rm TC}$ and $\SO(N)_{\rm TC}$ techni-color gauge groups, but
we restrict to techni-quarks in the fundamental representations of the TC group.

\item We consider techni-quark representations that can be embedded in SU(5)-unified models. 

\item 
We do not consider techni-scalars, that would generate a different set of TCb, and would allow to
realise  partial compositeness in a fundamental theory.

\end{enumerate}

\subsubsection*{The accidentally stable Dark Matter candidates}

This scenario has the following accidental symmetries that lead to  automatically stable composite DM candidates:
\begin{itemize}

\item{\bf Techni-baryon number}. The Lagrangian is accidentally symmetric under a $\U(1)_{\rm TB}$ global symmetry 
(sometimes broken by anomalies down to $\Z_2$) that rotates  the techni-quarks $\Q$ with the same phase. This guarantees the {\em stability of the lightest techni-baryon}.

\item{\bf Species Number}.
When the techni-quarks are in a reducible representation of the SM,
each phase rotation acting individually on a $\Q_i$ is an accidental techni-flavor
symmetry of the renormalizable Lagrangian. 
This leads to {\em stable techni-pions made of different species $\bar{\Q}_i \Q_j$}. TCb made
of different species can also be stable if their decay to TC$\pi$ is kinematically forbidden.

\item{\bf $G$-parity}. In models with electro-weak representations the Lagrangian can be invariant under a discrete symmetry known as $G$-parity \cite{hill},
that  acts on techni-quarks as \mbox{$\Q\to \exp (i \pi T^2)\Q^c$}.  In $\SU(N)_{\rm TC}$ theories $G$-parity acts on TC$\pi$ so that even (odd) isospin TC$\pi$ are even (odd) under $G$-parity. 
Standard Model states are $G$-parity even, so that the lightest $G$-parity odd TC$\pi$ is stable. This symmetry is broken by non-vanishing 
hypercharge.

\end{itemize}
We assume that, in a successful model, all stable particles must be good DM candidates.

\subsubsection*{Breaking of accidental symmetries}\label{sec:breaking}

The symmetries above can be violated by various effects. 

First, when the quantum numbers allow for
Yukawa interactions with the Higgs, this breaks  both species number and $G$-parity while preserving techni-baryon number.
States whose stability was insured by these broken symmetries will then decay  with specific patterns.
We assume that all allowed couplings are present and that
decays are fast enough that unstable particles are not relevant for dark matter. 


Second, species number and $G-$parity  can also be broken by dimension 5 operators,
\begin{equation}
\frac 1 M \bar{\Q} \Q H H\,,\qquad 
\frac{1}{{M}}
\bar{\Q} \sigma^{\mu\nu}\Q B_{\mu\nu} \,.
\end{equation}
The  lifetime of TC$\pi$ is  shorter than the age of the universe for $M < \bar M_{\rm Pl}\equiv 2.4\times 10^{18} $ GeV.


Third, techni-baryon  and species number can be broken by operators of dimension 6 or higher depending on the quantum numbers.
In the first case the lifetime is consistent with the present experimental bound from indirect searches~\cite{Sannino} 
\begin{equation}
\tau \sim   \frac{8\pi\,M^4}{M_{\rm DM}^5}\sim 10^{26}\,{\rm sec}\times
 \left( \frac {M}{\rm {\bar M}_{\rm Pl}}\right)^4  \left( \frac {100\,{\rm TeV}}{M_{\rm DM}}\right)^5  \circa{>}
 10^{25}\,{\rm sec} \frac{100\TeV}{M_{\rm DM}} 
\end{equation}
if $M$ is comparable to $\bar M_{\rm Pl}$ and $M_{\rm DM}\approx 100\TeV$.
For dimension 7 operators the scale $M$ must be larger than $\approx 10^{14}\,{\rm GeV}$. 

Any species number symmetry can also be broken by adding e.g.\ ad-hoc scalars with quantum numbers such that desired extra Yukawa couplings arise.

\medskip

The upshot is that techni-baryon number is more robust than species number or $G$-parity in the framework of vector-like confinement,
at least working within the standard assumptions of effective field theory. TCb are then the most promising dark matter candidate. 
We will focus mostly  on TCb dark matter in what follows.

\bigskip

The paper is structured as follows.
We identify successful DM models based on $\SU(N)_{\rm TC}$ in section~\ref{SU} 
and models based on $\SO(N)_{\rm TC}$ in section~\ref{SO}.
In section~\ref{spec} we discuss the effect of techni-quark masses and of the $\theta_{\rm TC}$ on the spectrum and the generation of Electric Dipole moments.
In section~\ref{pheno} we discuss the resulting phenomenology. Conclusions are given in section~\ref{end}. In the appendices we provide technical details of the
techni-baryon classification and we collect models that require higher dimensional operators.

\begin{table}[t]
\begin{center}
\begin{tabular}{|c|ccc|c|c|ccc|}
\hline 
$\SU(5)$ & $\SU(3)_c$ & $\SU(2)_L$ & U$(1)_Y$ & charge & name & $\Delta b_3$ & $\Delta b_2$ & $\Delta b_Y$ \\
\hline \hline
\rowcolor[cmyk]{0,0,0.2,0} $1$ & $1$ & $1$ & $0$ & $0$ & $N$ & 0 & 0 & 0 \\  \hline
\rowcolor[cmyk]{0,0.2,0,0.1} $ \bar{5}$ & $\bar{3}$ & $1$ & $1/3$ & $1/3$ & $D$ & 1/3 & 0 & 2/9 \\
\rowcolor[cmyk]{0,0.2,0,0.1}  & $1$ & $2$ & $-1/2$ & $0$,\,$-1$ & $L$ & 0 & 1/3 & 1/3 \\
\hline
\rowcolor[cmyk]{0.1,0,0.1,0} $10 $ & $\bar{3}$ & $1$ & $-2/3$ & $-2/3$ & $U$ & 1/3 & 0 & 8/9 \\
\rowcolor[cmyk]{0.1,0,0.1,0} & $1$ & $1$ & $1$ & $1$ & $E$ & 0 & 0 & 2/3 \\
\rowcolor[cmyk]{0.1,0,0.1,0} & $3$ & $2$ & $1/6$ & $2/3$,\,$-1/3$ & $Q$ & 2/3 & 1 & 1/9 \\
\hline
\rowcolor[cmyk]{0.2,0.0,0,0.0} $15 $ & $3$ & $2$ & $1/6$ & $2/3$,\,$-1/3$ & $Q$ & $2/3$ & $1$ & $1/9$ \\
\rowcolor[cmyk]{0.2,0.0,0,0.0} & $1$ & $3$ & $1$ & $0$,\,$1$,\,$2$ & $T$ & 0 & 4/3 & 2 \\
\rowcolor[cmyk]{0.2,0.0,0,0.0} & $6$ & $1$ & $-2/3$ & $-2/3$ & $S$ & 5/3 & 0 & 8/9 \\
 \hline
\rowcolor[cmyk]{0,0,0.2,0} $24$ & $1$ & $3$ & $0$ & $-1,\,0,\,1$ & $V$ & 0 & 4/3 & 0 \\
\rowcolor[cmyk]{0,0,0.2,0} & $8$ & $1$ & $0$ & $0$ & $G$ & 2 & 0 & 0 \\
\rowcolor[cmyk]{0,0,0.2,0} & $\bar{3}$ & $2$ & $5/6$ & $4/3$,\,$1/3$ & $X$ & 2/3 & 1 & 25/9 \\
\rowcolor[cmyk]{0,0,0.2,0} & $1$ & $1$ & $0$ & $0$ & $N$ &0 & 0 & 0 \\
\hline
\end{tabular}
\end{center}
\caption{\label{tab:unificaxion} \em  Techni-quarks are assumed to belong to fragments of $\SU(5)$ representations
(plus their conjugates for complex representations).
We give the SM decomposition, assign standard names used throughout the paper, and list the contributions  $\Delta b_i$  to the SM
$\beta$-function coefficients (to be multiplied by the multiplicity of the techni-color representation).}
\end{table}

\section{SU($N$)$_{\rm TC}$ Composite Dark Matter models} \label{SU}

In this section we consider an $\SU(N)_{\rm TC}$ techni-color group with $\NF$  techni-quarks in its fundamental representation.  
We assume that the dynamics is as in QCD: when techni-color interactions become strong, confinement takes place and
the global flavor symmetry $\SU(\NF)_L\otimes \SU(\NF)_R$ is spontaneously broken to the diagonal sub-group $\SU(\NF)$ producing $\NF^2-1$ Goldstone bosons in the adjoint representation of the unbroken group. We assume  the standard large $N$ scaling
\begin{equation}
\Lambda_{\rm TC}\sim \frac {4\pi}{\sqrt{N}} f\,,  \qquad m_B\sim N \Lambda_{\rm TC}
\end{equation}
where, to be definite, we denote with $\Lambda_{\rm TC}$ the mass of the lightest vector meson, with $f$  the Goldstone bosons decay constant, and with $m_B$ the techni-baryon mass.

We consider a model as viable from the point of view of Dark Matter phenomenology, provided that
all its stable states have no color, no charge and no hypercharge. 
This implies that dark matter should belong to a multiplet with integer isospin. As in weakly coupled
theories, the neutral component within an electroweak multiplet becomes the lightest component, 
with a calculable splitting, of order 100 MeV, induced by  electro-weak symmetry breaking \cite{MDM}.

We  analyzed these requirements using  the tools in appendix \ref{appendixA} and the package LieArt~\cite{lieart}. 
We assume an $\SU(5)$ unification scheme, so we select techni-quarks from components  of the simpler $\SU(5)$ representations listed in table~\ref{tab:unificaxion}. 
In general for a SM  representation  there are two inequivalent assignments of techni-quark quantum numbers:
\begin{equation}
R \equiv R_N \oplus \bar R_{\bar N},\quad \hbox{and}\qquad
\tilde R \equiv \bar R_N \oplus  R_{\bar N}
\label{vector}
\end{equation}
where $R_N$ and $\bar R_N$ transform in the fundamental of $\SU(N)_{\rm TC}$,
while $R_{\bar N}$ and $\bar R _{\bar N}$  in anti-fundamentals.
Since the $V$, $N$ and $G$ representations are real under the SM gauge group, one has $V=\tilde{V}$, $N=\tilde{N}$ and $G=\tilde{G}$.
For each SM representation, an unbroken species symmetry exists corresponding to a $\U(1)$ that rotates the (anti)fundamental of $\SU(N)_{\rm TC}$ 
with charge $+1$ ($-1$). Because of this accidental symmetry, TC$\pi$ made by different species are stable unless the symmetry is  broken e.g.\ by Yukawa couplings.  
Techni-baryon number, that guarantees the stability of the lightest TCb, is the sum of all species numbers.

It is convenient to classify models in the following way:

\begin{enumerate}
\item 

{\bf Golden-class} models, such that all stable states are acceptable DM candidates with just renormalizable interactions\footnote{The dimension-less models considered in~\cite{us} are a sub-set of these models, with the extra assumption of vanishing techni-quark masses.}.
Yukawa couplings are often needed in order to break  accidental symmetries, avoiding unwanted stable TC$\pi$.
All possible Yukawa couplings among the $\SU(5)$ fragments are:
\beq \label{eq:SUyuk}
HL  (\tilde{E}\hbox{ or }\tilde{T}\hbox{ or }N\hbox{ or }V),\qquad
HQ  (\tilde{D}\hbox{ or }\tilde{U}),\qquad
HD  {X},\qquad
\eeq
as well as similar interactions with $H\leftrightarrow H^\dagger$ or $x\leftrightarrow \tilde{x}$ where $x$ denotes all techni-quarks.

\item
{\bf Silver-class} models where non-renormalizable interactions or ad-hoc extra particles
are introduced in order to break accidental symmetries that lead to unwanted stable particles\footnote{
Higher dimensional operators violate flavour in general. Assuming that the scale suppressing these operators is
around the GUT or Planck scale, as required for baryon violating operators, this does not  lead to phenomenological problems.} 

\item Models with no DM candidates.

\end{enumerate}

An important restriction on the techni-quark content arises from the requirement that  $\SU(N)_{\rm TC}$ with $\NF$ flavors
of techni-quarks  (e.g.\ a singlet $N$ contributes as $\NF=1$)
 is asymptotically free.  Defining the gauge $\beta$-function coefficients as
 $ d\alpha_i^{-1} / {d \log Q} =- {b_i}/{2\pi}$
we have  
\beq b_{\rm TC} = - {\frac{11}{3} N} + \frac23 \NF < 0.\eeq
Furthermore we demand that the SM gauge couplings do not develop Landau poles below the Planck scale:
\beq 
b_3 = -7 +\Delta b_3  \circa{<}3,\qquad
b_2 = -\frac{19}{6}+\Delta b_2\circa{<} 6.5 ,\qquad
b_Y = \frac{41}{6}+\Delta b_Y\circa{<}18.
\eeq
where the numerical factors have been computed assuming $\Lambda_{\rm TC}\sim $100 TeV, motivated by DM as a thermal relic, see section \ref{pheno}.
Colored techni-quarks such as $U$ or $D$ contribute as $\Delta b_3 = 2N/3$, while a $G$ state gives $\Delta b_3=4N$.
The weak doublet $L$ contributes as $\Delta b_2=2N/3$, while for the weak triplet $V$ we have $\Delta b_2 = 8N/3$.
Finally $\Delta b_Y = \frac23  \sum_R {\rm dim}(R) Y_R^2$ (e.g.\ a singlet $E$ contributes as $\Delta b_Y = 4N/3$).
The contributions $\Delta b_{2,3,Y}$ are summed over techni-quarks, and the constant terms in the $\beta$-function coefficients $b_{2,3,Y}$ are the SM contributions.

Summarising, the constraints on the techni-quark content are:
\beq \NF < {\frac{11}{2} N} ,\qquad \Delta b_3 \circa{<} 10,\qquad
\Delta b_2 \circa{<}10,\qquad
\Delta b_Y \circa{<}11.
\label{eq:Nbound}
\eeq
This implies that one weak triplet $V$ is allowed by the constraint on $\Delta b_2$ for $N=3$ techni-colors but not for $N\ge 4$.
Models that contain the techni-quark $G$, $S$, $X$ are not allowed, not even for $N=3$, because of $\Delta b_3$ or $\Delta b_Y$.

\bigskip

\subsection{Techni-pions and techni-baryons of SU($N$)$_{\rm TC}$} \label{sec:SUN-pions-baryons}
Techni-pions are $\Q\bar \Q$ states in the adjoint representations of $\SU(\NF)$ under the unbroken techni-flavor symmetry. 
Their decomposition under the SM group is  given by
\begin{equation}\label{RRbar}
{\rm Adj}_{\SU(\NF)}= \left[\sum_{i=1}^{N_S} R_i \right]\otimes \left[\sum_{i=1}^{N_S}  \bar{R}_i \right] \ominus 1
\end{equation}
where the sum runs over the $N_S$ species (e.g.\ a model with $\Psi = L \oplus N$ techni-quarks has $N_S=2$ species and $\NF = 2+1$ techni-flavors). 
SM gauge interaction generate a positive contribution to TC$\pi$ masses that can be estimated as
\begin{equation}
\Delta_{\rm gauge} m_{{\rm TC}\pi}^2\sim  \frac {g^2}{(4\pi)^2}  \Lambda_{{\rm TC}}^2.
\label{mpigauge}
\end{equation}
The $N_S-1$ singlets under the SM gauge group do not acquire mass from gauge interactions.
In our previous study~\cite{us} we assumed vanishing techni-quark masses, such that these singlet TC$\pi$ 
were massless in absence of Yukawa interactions, and thereby experimentally excluded because of their axion-like coupling to SM vectors.
Here we allow for techni-quark masses, such that the singlets become massive avoiding phenomenological problems.  
The contribution from techni-quark masses to TC$\pi$ masses scales as
\beq
\Delta_{\rm mass} m_{{\rm TC}\pi}^2\sim m_\Q    \Lambda_{{\rm TC}} \,
\label{mpimass}
\eeq
and can be described using chiral Lagrangian techniques. 

Techni-pions can be stable because of $G$-parity or  species number if they are made by different species. 
For example in QCD,  the charged pion $\pi^+$ decays because species number is broken by weak interactions,
while $G-$parity is broken by hypercharge allowing $\pi_0$ to decay through the anomaly. 
Among our representations, only the weak triplet $V$ is symmetric under $G$-parity leading to stable TC$\pi$.

\bigskip

TCb are techni-color singlets constructed with $N$ techni-quarks. 
They are fermions for $N$ odd and bosons for $N$ even, leading to vastly different dark matter phenomenology. 
The SM quantum numbers of TCb multiplets are determined by group theory:
the TCb fill representations of the unbroken $\SU(\NF)$ global techni-flavor symmetry that can be decomposed under the SM. 
TCb wave-function is totally antisymmetric  in techni-color. Furthermore,  one can argue that the lighter TCb have the smallest possible spin, 
and the lowest possible angular momentum (fully symmetric $s$-wave function in space).
Due to Fermi statistics, this implies that TCb must be fully symmetric in spin and techni-flavour. 
This determines the representation of the lighter TCb under the unbroken global techni-flavor symmetry
corresponding to a Young tableau with two rows with $N/2$ boxes ($N$ even) or two  
rows with $(N+1)/2$ and  $(N-1)/2$ boxes ($N$ odd) and also the spin. 
Explicitly for $N=3,4,5$ they are,
\begin{equation}
\hbox{lighter TCb} = \left\{\begin{array}{ll}
{\tiny \Yvcentermath1  \yng(2,1)} & \hbox{for $N=3$}\\
{\tiny \Yvcentermath1  \yng(2,2) } & \hbox{for $N=4$}\\
{\tiny \Yvcentermath1\yng(3,2) } & \hbox{for $N=5$} \,.
\end{array}\right.
\label{youngtableauxlight}
\end{equation}
A fully symmetric representation is obtained by a tensor product of each techni-flavor representation with an identical spin representation:
for even (odd) $N$ we obtain spin-0 (spin 1/2) DM.
The case $\NF=1$ is special because flavour cannot be anti-symmetrized, TCb have spin $N/2$.
The heavier TCb (analog of the decuplet in QCD)  transform instead in the following representations
\begin{equation}
\hbox{heavier TCb} = \left\{\begin{array}{ll}
{\tiny \Yvcentermath1   \yng(3)} & \hbox{for $N=3$}\\
{\tiny \Yvcentermath1  \yng(3,1)\oplus \yng(4) } & \hbox{for $N=4$}\\
{\tiny \Yvcentermath1  \yng(4,1)\oplus \yng(5) } & \hbox{for $N=5$}\\
\end{array}\right.
\end{equation}
and have higher spin described by an identical spin representation. 
The mass difference between the heavier and the lighter TCb is expected of order $\Lambda_{\rm TC}$.

Heavier TCb usually decay into a lighter TCb and TC$\pi$;
however some heavier TCb could be accidentally stable due to species number
if they are the lightest states with TCb and species number. This can happen for techni-quark masses 
comparable to $\Lambda_{\rm TC}$. An analog exists in QCD 
where, in absence of the weak interactions, the lightest strange baryon ($\Lambda$, with quark content $uds$)
would be stable because its decay to kaons and nucleons  is not kinematically allowed.
Furthermore, the spin 3/2 baryon $\Omega^-(1672)$ (quark content $sss$)
cannot decay to $\Xi^0K^-$ through strong interactions: its decay is allowed only by
strangeness-violating weak interactions. 

\bigskip

\begin{figure}[t]
$$\includegraphics[width=0.8\textwidth]{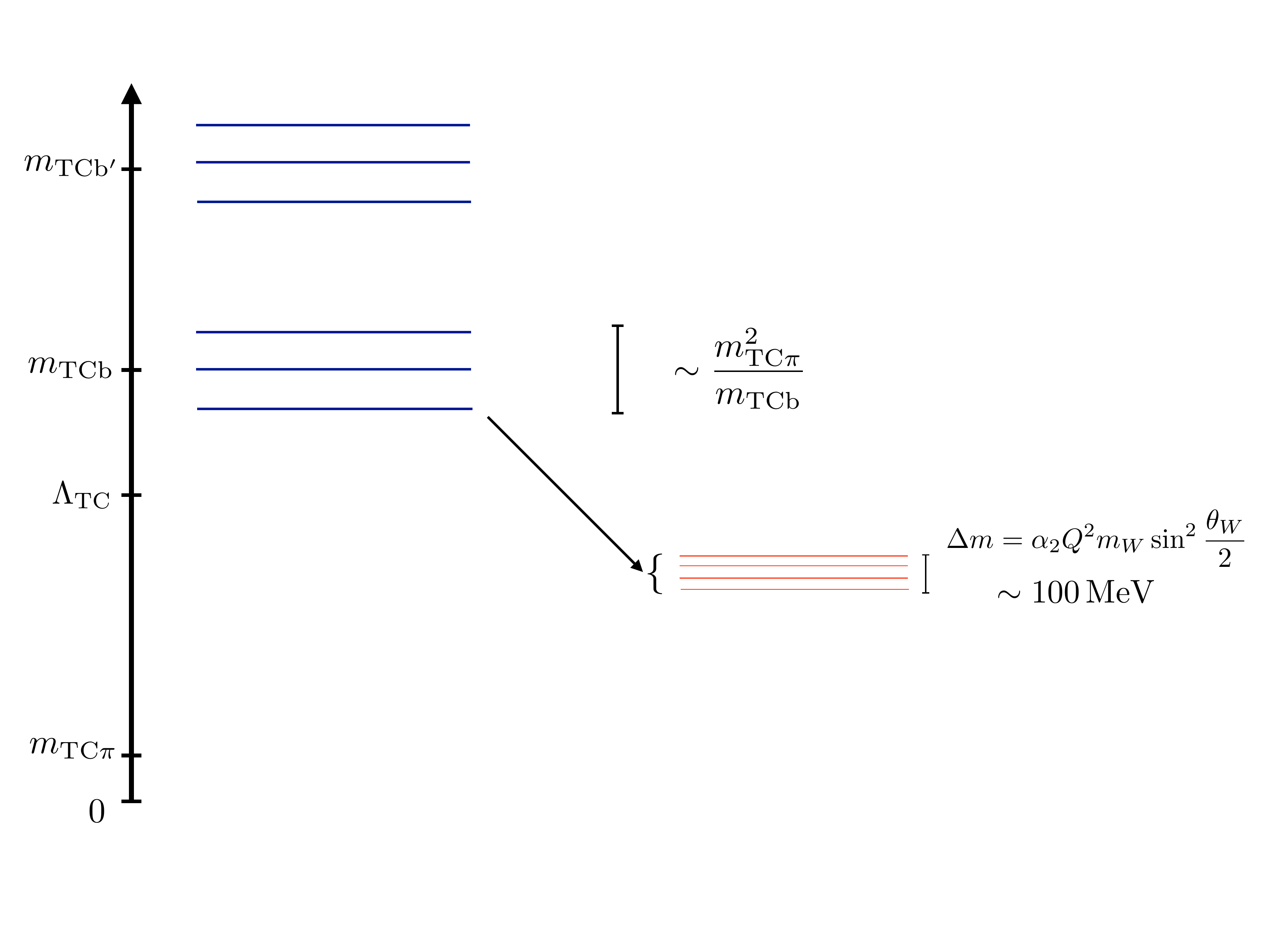}$$
\caption{\em Spectrum of techni-color DM models. Splitting between techni-flavor multiplets is of order the dynamical scale 
$\Lambda_{{\rm TC}}$ splitting between different SM representation $\Lambda_{{\rm TC}}/100$ or larger and hyperfine 
splitting from electro-weak symmetry breaking of order $100\MeV$.
\label{fig:spectrum}}
\end{figure}

TCb flavour multiplets are split by SM gauge interactions, by techni-quark masses
and possibly by techni-quark Yukawa interactions and by higher dimensional operators (that we neglect).
While for the TC$\pi$ one can argue that in the  limit of zero techni-quark masses the lightest multiplets are
those with the smallest charge under the SM gauge group,  the same sentence is not rigorously proved for TCb. 
Indeed, while the long distance gauge contribution to the energy of charged fields 
is  proportional to their total charge, the short distance contribution is difficult to estimate. 
Experience with electromagnetic splitting of baryons in QCD
hints however to the fact that the lightest states are indeed the ones with smaller charge.
This is what we will assume in the following. We estimate,
\begin{equation}
\Delta_{\rm mass} m_B\sim m_\Q\,,\qquad
\Delta_{\rm gauge} m_B\sim \frac {g^2}{(4\pi)^2} \Lambda_{{\rm TC}}.
\end{equation}

Finally, the breaking of the electro-weak symmetry  induces calculable splittings within the components of each electro-weak multiplet (of order 100 MeV), 
with the result that the component with smallest electric charge  is the lightest state. The spectrum of the theory is illustrated in fig.~\ref{fig:spectrum}.

\bigskip

\subsection{SU($N$)$_{\rm TC}$ golden-class models} \label{sec:SU3-models}
In this section we present the golden-class models for $\SU(N)_{\rm TC}$ strong interactions.
The models are obtained scanning over techni-quarks made by combinations of the $\SU(5)$ fragments of table~\ref{tab:unificaxion}. 
excluding models that lead to sub-Planckian Landau poles for $g_Y$, $g_2$ or $g_3$. We require that the lightest stable TCb has no color, no hypercharge, and integer isospin. 
For example, for $N=3$, the possible DM candidates are made of the following techni-quarks:
\beq 
LLE, ~DDU, ~ EU\tilde D, ~ QQ\tilde{D} ~DLQ,  ~ UQ\tilde L, ~ V x\tilde{x},~
\label{eq:Blist}
\eeq
where $x$ denotes any techni-quark, any $E$ can be substituted by a $T$,  any $V$ can be substituted by a $N$. By replacing all techni-quarks with their tilded counterparts one obtains equivalent descriptions of the same models.

However, if species number is conserved,
most of the models that can give rise to such TCb DM candidates
also lead to extra stable TC$\pi$ with $Y\neq 0$ or color, that are thereby excluded by DM direct searches
(unless their thermal abundance is small enough).
In the context of renormalizable golden-class models, 
Yukawa couplings to the Higgs doublet determine the accidental symmetries.
For example, a Yukawa coupling to the Higgs boson is allowed by gauge quantum numbers in a model containing the techni-quarks $\Q = L\oplus\tilde{E}$. The Yukawa coupling $H L \tilde{E}$ breaks the unwanted species number.
On the contrary, no Yukawa coupling is allowed in a model
with $\Q = L\oplus E$ that would lead to the first TCb in eq.\eq{Blist}. 
In appendix~\ref{sec:silvermodels} we present a list of  silver-class models (limited for simplicity to $N=3,4$ and to the case of 1 or 2 species)
where extra effects (non-renormalizable interactions or other particles) are needed to break unwanted symmetries.

\begin{table}[p]\small
$$\hspace{-0ex}\begin{array}{ccc|cc|c}
\multicolumn{1}{c}{\hbox{$\SU(N)$ techni-color.}} & \hbox{Yukawa}&\hbox{Allowed}
&\hbox{Techni-}& \multicolumn{1}{c|}{\hbox{Techni-}} &   \\  
\multicolumn{1}{c}{\hbox{Techni-quarks}}&\hbox{couplings}&N &
 \hbox{pions} & \hbox{baryons} &\hbox{under} \\
 \hline
\rowcolor[cmyk]{0,0,0.2,0.1} \multicolumn{1}{c}{\NF=3} & &&  8 & 8 , \bar 6, \ldots \hbox{ for } N=3,4,\ldots&  \hbox{$\SU(3)_{\rm TF}$}  \\ 
\rowcolor[cmyk]{0,0,0.2,0}  \Q =  V   &0&3& \hbox{3} & VVV=3  & \hbox{$\SU(2)_L$}\\
 \rowcolor[cmyk]{0,0,0.2,0}  \Q =  N\oplus L   &\hbox{1}&3,..,14& \hbox{unstable} & N^{N*}=1  & \hbox{$\SU(2)_L$}\\
 \hline
\rowcolor[cmyk]{0,0,0.2,0.1}
\multicolumn{1}{c}{\NF=4} && & 15& \overline{20}, 20',\ldots & \hbox{$\SU(4)_{\rm TF}$}\\
\rowcolor[cmyk]{0,0,0.2,0}  \Q=V\oplus N  &\hbox{0}&3& 3 \times 3 & VVV, VNN=3 ,~ VVN=1 & \hbox{$\SU(2)_L$}\\
 \rowcolor[cmyk]{0,0,0.2,0} \Q =  N\oplus L\oplus \tilde E   &\hbox{2}&3,4,5 & \hbox{unstable} & N^{N*}=1 & \hbox{$\SU(2)_L$}\\
 \hline
\rowcolor[cmyk]{0,0,0.2,0.1} 
\multicolumn{1}{c}{\NF=5} & & & 24 & { \overline{40}},\overline{50} & \hbox{$\SU(5)_{\rm TF}$}\\
\rowcolor[cmyk]{0,0,0.2,0}
\Q=V \oplus L &\hbox{1}&3& \hbox{unstable} & VVV=3 & \hbox{$\SU(2)_L$} \\
\rowcolor[cmyk]{0,0,0.2,0}\Q=N \oplus L\oplus\tilde{L}&2&3& \hbox{unstable} &  NL\tilde{L}=1 & \hbox{$\SU(2)_L$} \\
\rowcolor[cmyk]{0,0,0.2,0}=&2&4& \hbox{unstable} &  NNL\tilde{L}, L\tilde{L}L\tilde{L}=1& \hbox{$\SU(2)_L$} \\
\hline
\rowcolor[cmyk]{0,0,0.2,0.1} 
\multicolumn{1}{c}{\NF=6} & && 35 & 70, \overline{105'}  & \hbox{$\SU(6)_{\rm TF}$}\\
\rowcolor[cmyk]{0,0,0.2,0}\Q=V \oplus L\oplus N  &\hbox{2}&3& \hbox{unstable} & VVV,VNN=3 , VVN=1   & \hbox{$\SU(2)_L$} \\
\rowcolor[cmyk]{0,0,0.2,0}\Q=V \oplus L\oplus \tilde E  &\hbox{2}& 3&\hbox{unstable} & VVV=3   & \hbox{$\SU(2)_L$} \\
\rowcolor[cmyk]{0,0,0.2,0}\Q=N \oplus L  \oplus \tilde L \oplus \tilde E  &\hbox{3}&3& \hbox{unstable} & NL\tilde{L},  \tilde L \tilde L \tilde E=1  & \hbox{$\SU(2)_L$} \\
\rowcolor[cmyk]{0,0,0.2,0}=  &\hbox{3}&4& \hbox{unstable} & NNL\tilde{L}, L\tilde{L}L\tilde{L}, N\tilde E \tilde L \tilde L =1   & \hbox{$\SU(2)_L$} \\  \hline
\rowcolor[cmyk]{0,0,0.2,0.1} \multicolumn{1}{c}{\NF=7} &&& 48 & 112 & \hbox{$\SU(7)_{\rm TF}$}\\
\rowcolor[cmyk]{0,0,0.2,0}
\Q=L \oplus \tilde L \oplus E \oplus \tilde E \oplus N  &\hbox{4}&3& \hbox{unstable} & LLE,\tilde L \tilde L \tilde E,  L \tilde L N,  E \tilde E N=1  & \hbox{$\SU(2)_L$} \\

\rowcolor[cmyk]{0,0,0.2,0}
\Q= N \oplus L \oplus \tilde E \oplus V  &\hbox{3}&3& \hbox{unstable} &  VVV, VNN=3 ,~ VVN=1 & \hbox{$\SU(2)_L$} \\

\hline

\rowcolor[cmyk]{0,0.2,0,0.1}
\multicolumn{1}{c}{\NF=9} &&& 80 & 240 & \hbox{$\SU(9)_{\rm TF}$}\\
\rowcolor[cmyk]{0,0.2,0,0}\Q=Q \oplus \tilde{D}&\hbox{1}&3& \hbox{unstable} & QQ\tilde{D}=1 & \hbox{$\SU(2)_L$} \\
\hline
\rowcolor[cmyk]{0,0.2,0,0.1}
\multicolumn{1}{c}{\NF=12} &&& 143 & 572 & \hbox{$\SU(12)_{\rm TF}$}\\
\rowcolor[cmyk]{0,0.2,0,0}\Q=Q \oplus \tilde{D}\oplus\tilde{U}  
&\hbox{2}&3& \hbox{unstable} & QQ\tilde{D},  \tilde D\tilde D\tilde U=1 & \hbox{$\SU(2)_L$} \\
\hline
\end{array}
$$
\caption{\em Golden-class models with $\SU(N)_{\rm TC}$ techni-color that give viable TCb and/or
TC$\pi$ Dark Matter candidates with $Q=Y=0$, starting from techni-quarks
coming from $\SU(5)$ fragments listed in table~\ref{tab:unificaxion}.
The darker rows give the  techni-flavour content of the lightest TCb and TC$\pi$ considering only masses induced by techni-color interactions.
The lighter rows  show the viable models, the number of Yukawa interactions,
and the $\SU(2)_L$ content of the stable TC$\pi$ and the stable TCb,
assuming that the lighter component is the one with the least SM charge.
A $^*$ denotes a higher spin TCb.
\label{dmsummaryfinal}}
\end{table}

\medskip

The list of $\SU(N)_{\rm TC}$ golden-class models presented below is summarized in table~\ref{dmsummaryfinal}\footnote{We do not consider models that contain SM representations with multiplicity as these do not lead to new DM candidates. In some cases however this might change the spin of the lightest TCb.}.
We start the description of golden-class models from models that only involve color-less techni-quarks.

The simplest model contains the singlet $N$ as the only techni-quark,
such that the lightest DM TCb has spin $N/2$. Interactions with SM particles
arise only adding extra states, as described below.

\subsubsection*{a) SU($N$)$_{\rm TC}$ model $\Q=V$}
The model has a single specie of techni-quarks:
a triplet with zero hypercharge in the adjoint of $\SU(2)_L$, such that $\NF=3$.
No Yukawa coupling is allowed.
If $N\ge 4$ the $g_2$ gauge coupling becomes non-perturbative
below the Planck scale.  Thereby this model is only allowed
for $N=3$.
Both TCb and TC$\pi$ 
lie in the 8 of $\SU(3)_{\rm TF}$, that decomposes as
\beq 8 = 3_0\oplus 5_0\qquad
\hbox{under $\SU(2)_L\otimes{\rm U}(1)_Y$}.\eeq
The TC$\pi$ triplet is stable because of $G$-parity, and the TCb triplet is stable because of techni-baryon number. These are good DM candidates. 
This model has been already presented in~\cite{us}.

\subsubsection*{b) SU($N$)$_{\rm TC}$ model $\Q=V\oplus N$}
The previous model can be simply extended to $N_S=2$ techni-quarks by adding an $N$ (SM gauge singlet)
such that $\NF=4$. Again, no Yukawa coupling is allowed and the model can be considered only for $N=3$ because of sub-Planckian Landau poles.
TC$\pi$ lie in the 15 of $\SU(4)_{\rm TF}$ that decomposes as
\beq \hbox{TC$\pi$ : } 15 = 1_0 \oplus 3 \times 3_0 \oplus 5_0
\qquad
\hbox{under $\SU(2)_L\otimes{\rm U}(1)_Y$}.\eeq
The three triplets are stable because of species number and because of $G$-parity.
 The lighter TCb lie in the $\overline{20}$ representation of $\SU(4)_{\rm TF}$ that decomposes 
as
\beq \hbox{TCb : } \overline{20} = 1_0 \oplus 3 \times 3_0 \oplus 2\times 5_0
\qquad
\hbox{under $\SU(2)_L\otimes{\rm U}(1)_Y$}.\eeq
The lightest TCb is a stable DM candidate, and its identity depends on the techni-quark masses. For $m_V \ll m_N$, the triplet $3_0~ (VVV)$ is expected to be the lightest.
For $m_N \ll m_V \circa{>}\Lambda_{\rm TC}$ the extra TCb $NNN^*$ (denoted with a $^*$ and not included in the  list above because it has spin 3/2)
could become the stable DM candidate; at the same time the $\SU(4)_{\rm TF}$ classification breaks down.

\subsubsection*{c) SU($N$)$_{\rm TC}$ models $\Q=N \oplus L$ and $\Q=N \oplus L \oplus \tilde E$}
In both models, enough Yukawa couplings are allowed such that only techni-baryon number is conserved and all TC$\pi$ are unstable. 
For the $N \oplus L$ ($\NF=3$) and the $N \oplus L \oplus \tilde E$ ($\NF=4$) models respectively, these are:
\beq\label{eq:TCpiNL}
\begin{array}{l}
\hbox{TC$\pi$ : } 8 = 1_0 \oplus 2_{\pm 1/2} \oplus 3_0 \\
\hbox{TC$\pi$ : } 15 = 1_{2 \times 0,\pm 1} \oplus 2 \times 2_{\pm 1/2} \oplus 3_0
\end{array}\qquad
\hbox{under $\SU(2)_L\otimes{\rm U}(1)_Y$}.
\eeq 
For $N=3$, the spin 1/2 TCb do not contain any DM candidate, for example in the
the $N \oplus L$ model they are
\beq
\label{N+L-TCb}
\hbox{TCb : } 8 = 1_{-1} \oplus 2_{-1/2,-3/2} \oplus 3_{-1} \qquad
\hbox{under $\SU(2)_L\otimes{\rm U}(1)_Y$}.
\eeq 
The DM candidate is the singlet $NNN^*$, which only exists with spin 3/2.
Thereby these models are viable only as long as the techni-quark masses $m_L$ and $m_{\tilde E}$ are 
of order $\Lambda_{\rm TC}$ and large enough that $NNN^*$ is the lightest TCb.
This state lies in the 10 of $\SU(3)_{\rm TF}$ in the $N\oplus L$ model 
\beq
\hbox{TCb* : } 10 = 1_0 \oplus 2_{-1/2} \oplus 3_{-1} \oplus 4_{-3/2} \qquad \hbox{under $\SU(2)_L\otimes{\rm U}(1)_Y$}
\eeq
and in the $\overline{20}''$ of $\SU(4)_{\rm TF}$ in the $N\oplus L \oplus \tilde E$ model.

The same is true for $N=4$, where the only DM candidate is the singlet $NNNN^*$ that lies in the completely symmetric spin 2 representation $\tiny \yng(4)$. In the $N \oplus L$ model, this representation decomposes as
\begin{eqnarray}
\hbox{TCb$^*$ : } 15' &=& 1_0 \oplus 2_{-1/2} \oplus 3_{-1} \oplus 4_{-3/2} \oplus 5_{-2} \qquad \hbox{under $\SU(2)_L\otimes{\rm U}(1)_Y$}.
\end{eqnarray}
The $N \oplus L$ model is allowed by perturbativity constraints up to $N=14$, while the $N \oplus L \oplus \tilde E$ is allowed up to $N=5$ (with increasing spin of the DM candidate).

\subsubsection*{d) SU($N$)$_{\rm TC}$ models $\Q = V\oplus L$ and $V \oplus L \oplus \tilde E$}

Other possible extensions of the first model are $\Q = V\oplus L$ and $V \oplus L \oplus \tilde E$. 
A possible problem of these models is that, even for $N=3$, the  $\SU(2)_L$ gauge coupling becomes non perturbative  around $10^{17}\GeV$.
In view of the Yukawa couplings $V  L H$, $\tilde E L H$,  all TC$\pi$ are unstable and given by
\beq 
\hbox{TC$\pi$ : } 24 = 1_0 \oplus 2_{\pm 1/2} \oplus 2 \times 3_0 \oplus 4_{\pm 1/2} \oplus 5_0 \qquad
\hbox{under $\SU(2)_L\otimes{\rm U}(1)_Y$} \,,
\label{LVpions}
\eeq
in the $V \oplus L$ model, and by
\beq 
\hbox{TC$\pi$ : } 35 = 2 \times 1_0 \oplus 2 \times 2_{\pm 1/2} \oplus 3_{2 \times 0,\pm 1} \oplus 4_{\pm 1/2} \oplus 5_0 \qquad
\hbox{under $\SU(2)_L\otimes{\rm U}(1)_Y$} \,,
\eeq
in the $V \oplus L \oplus \tilde E$ model.

In both models the TCb DM candidate is the $VVV$ state that forms a weak triplet
as in the $\Q = V$ model: the  extra techni-quarks $L$ (and possibly $\tilde E$)
does not lead to any extra DM candidates and play a minor role provided that they are heavy enough.
In the $V \oplus L$ model, the lightest TCb multiplet is a $\overline{40}$ of $\SU(5)_{\rm TF}$ that decomposes as:
\beq \hbox{TCb : } \overline{40} = 1_{-1} \oplus 2_{2 \times (-1/2),-3/2} \oplus 3_{0, 2 \times (-1)} \oplus 2 \times 4_{-1/2} \oplus 5_{0,-1} \oplus 6_{-1/2}\eeq
under $\SU(2)_L\otimes{\rm U}(1)_Y$.

\subsubsection*{e) SU($N$)$_{\rm TC}$ models $\Q = V\oplus N\oplus L$ and $\Q = V \oplus  N \oplus L \oplus \tilde E$}
As in the previous models, sub-Planckian Landau poles are avoided only for $N=3$
(where $g_2$ becomes non perturbative around $10^{17} \GeV$).
Since $L$ and $\tilde E$ cannot enter in an hypercharge-less TCb, the DM candidates are the same of the $V \oplus N$ model.   Unlike in the $V\oplus N$ model, the Yukawa couplings $V L H$, $N L H$ and $L \tilde E H$ 
break all species number symmetries, such that all 
TC$\pi$ are unstable. 
In the $V \oplus N \oplus L$ model ($\NF=6$), the TC$\pi$ are
\beq 
\hbox{TC$\pi$ : } 35 = 2 \times 1_0 \oplus 2 \times 2_{\pm 1/2} \oplus 4 \times 3_0 \oplus 4_{\pm 1/2} \oplus 5_0 \qquad
\hbox{under $\SU(2)_L\otimes{\rm U}(1)_Y$} \,.
\eeq
In the $V \oplus N \oplus L \oplus \tilde E$ model ($\NF=7$), the list extends to 
\beq 
\hbox{TC$\pi$ : } 48 = 1_{3 \times 0,\pm 1} \oplus 3 \times 2_{\pm 1/2} \oplus 3_{4 \times 0,\pm 1} \oplus 4_{\pm 1/2} \oplus 5_0 \qquad
\hbox{under $\SU(2)_L\otimes{\rm U}(1)_Y$} \,.
\eeq

\subsubsection*{f) SU($N$)$_{\rm TC}$ model $\Q=N \oplus L \oplus \tilde L$}
The model allows two Yukawa couplings
($N L H$, $N \tilde L H$) such that there are no stable TC$\pi$
and allows for DM TCb candidates not present in the previous models.
The unstable TC$\pi$ are:
\beq \label{eq:TCpiNLL}
\hbox{TC$\pi$ : } 24 = 1_{2 \times 0,\pm 1} \oplus 2\times 2_{\pm 1/2} \oplus 3_{2 \times 0,\pm 1} \qquad
\hbox{under $\SU(2)_L\otimes{\rm U}(1)_Y$} \,. \eeq
Sub-Planckian Landau poles are avoided for $N\le 7$.  Here we discuss the TCb DM candidates for $N=3,4$.

For $N=3$, the lighter TCb fill a $\overline{40}$ of $\SU(5)_{\rm TF}$ that decomposes as
\beq
\label{N+L+tL-TCb}
\hbox{TCb : } \overline{40} = 1_{2 \times 0,\pm 1} \oplus 2_{3 \times (\pm 1/2),\pm 3/2} \oplus 3_{2 \times 0,\pm 1} \oplus 4_{\pm 1/2}
\eeq
\hbox{under $\SU(2)_L\otimes{\rm U}(1)_Y$}, so that the TCb DM candidates are singlets made of $N \tilde L L$.  

For $N=4$, the lighter TCb are
\beq 
\label{TCbNLL}
\hbox{TCb : } \overline{50} = 1_{3 \times 0,\pm 1,\pm 2} \oplus 2_{2 \times (\pm 1/2),\pm 3/2} \oplus 3_{2 \times 0, 2 \times (\pm 1)} \oplus 4_{\pm 1/2} \oplus 5_0 \,,\eeq
under $\SU(2)_L\otimes{\rm U}(1)_Y$.
The TCb DM candidates are singlets made of $LL \tilde L \tilde L$ and $L \tilde L NN$.

\subsubsection*{g) SU($N$)$_{\rm TC}$ model $\Q = N\oplus L \oplus \tilde L \oplus \tilde E$}
This is a non trivial extension of the previous model, with one more Yukawa coupling allowed ($L \tilde E H$), so that there are no stable TC$\pi$. 
The model is allowed only for $N=3,4$, since for greater values of $N$ the coupling $g_Y$ develops a sub-Planckian Landau pole.
The unstable TC$\pi$ can be listed as:
\beq \label{eq:TCpiNLLE}
\hbox{TC$\pi$ : } 35 = 1_{3 \times 0, 2 \times (\pm 1)} \oplus 2_{3 \times (\pm 1/2),\pm 3/2} \oplus 3_{2 \times 0,\pm 1} \qquad
\hbox{under $\SU(2)_L\otimes{\rm U}(1)_Y$} \,. \eeq
This model gives a TCb DM candidate not present in the previous models: $\tilde L \tilde L \tilde E$ and $N \tilde L \tilde L \tilde E$ for $N=3$ and $N=4$ respectively.

For $N=3$, the lightest multiplet of TCb decomposes under $\SU(2)_L\otimes{\rm U}(1)_Y$ as 
\begin{eqnarray}
\hbox{TCb : } 70 &=& \hbox{TCb}_{N\oplus L \oplus \tilde L} \oplus 1_{0, 3 \times (-1), 2 \times (-2)} \oplus 2_{2 \times (-1/2), 3 \times (-3/2),(-5/2)} 
\oplus 3_{0, 2 \times (-1),-2} \,,
\end{eqnarray}
where $\hbox{TCb}_{N\oplus L \oplus \tilde L}$ is defined in eq.~\eqref{N+L+tL-TCb}. 
For $N=4$ we get
\begin{eqnarray}
\hbox{TCb : } \overline{105'} &=& \hbox{TCb}_{N \oplus L \oplus \tilde L} \oplus 1_{0,2 \times (-1),3 \times (-2)} \oplus 2_{1/2, 3 \times (-1/2), 4 \times (-3/2), 2 \times (-5/2)} \notag \\
&&\oplus 3_{0, 3 \times (-1), 2 \times (-2),-3} \oplus 4_{-1/2,-3/2} 
\end{eqnarray}
where now $\hbox{TCb}_{N \oplus L \oplus \tilde L}$ refers to eq.~\eqref{TCbNLL}. 
In each case, besides the TCb DM candidates of the $N \oplus L \oplus \tilde L$ model, there a singlet DM candidate made of $\tilde L \tilde L \tilde E$ or $N \tilde L \tilde L \tilde E$.



\subsubsection*{h) SU($N$)$_{\rm TC}$ model $\Q = L \oplus \tilde L \oplus E \oplus \tilde E \oplus N$} 
The model  has  $\NF=7$ and for $N=3$ gives $\Delta b_Y=12$ so that hypercharge has a Landau pole around the Planck scale, so that it cannot be extended to $N>3$.
Thanks to the presence of $N$, 
it allows for 4 Yukawa couplings ($L \tilde E H$, $\tilde L E H$,  $LNH$, $\tilde{L}NH$) that break all species number symmetries.
The unstable TC$\pi$ are:
\beq
\hbox{TC$\pi$ : } 48 = 1_{4 \times 0, 3 \times (\pm 1),\pm 2} \oplus 2_{4 \times (\pm 1/2), 2 \times (\pm 3/2)} \oplus 3_{2 \times 0,\pm 1}
\eeq
\hbox{under $\SU(2)_L\otimes{\rm U}(1)_Y$}.
The lightest TCb fill a $112$ of $\SU(7)_{\rm TF}$, that decomposes as 
\begin{eqnarray}
\hbox{TCb : } 112 &=& \hbox{TCb}_{N\oplus L \oplus \tilde L} \oplus 1_{4 \times (0, \pm 1), 2 \times (\pm 2)} \oplus 2_{4 \times (\pm 1/2), 3 \times(\pm 3/2), \pm 5/2} \oplus 3_{2 \times (0,\pm 1),\pm 2} \,,
\end{eqnarray}
under $\SU(2)_L \otimes {\rm U}(1)_Y$.  
The TCb DM candidates are those of the $N \oplus L \oplus \tilde L$ model, defined in eq.~\eqref{N+L+tL-TCb}, plus the singlets $LLE$, $\tilde L \tilde L \tilde E$ and $E \tilde E N$.

\bigskip

We next consider models with coloured techni-quarks.

\subsubsection*{i) SU($N$)$_{\rm TC}$ model $\Q = Q\oplus \tilde D$}
The simplest golden-class model with colored techni-quarks is
%
$\Q=Q\oplus \tilde D$, that is allowed for $N=3,4$ and gives a DM candidate only for $N=3$.
The model has $\NF=6$ and does not lead to unwanted stable states because species number is broken by the  Yukawa coupling $Q\tilde{D} H$.

The model predicts a set of unstable TC$\pi$ 
 in the $80$ representation of $\SU(9)_{\rm TF}$,  
that decomposes under the SM gauge group $\SU(3)_c \otimes \SU(2)_L \otimes {\rm U}(1)_Y$ as
\beq \hbox{TC$\pi$ : } 80=
(1,1)_0 \oplus (1,2)_{\pm 1/2}  \oplus (1,3)_0 \oplus 2(8,1)_0 \oplus (8,2)_{\pm 1/2}  \oplus (8,3)_0.\eeq
For $N=3$ the multiplet of lighter TCb fills a $240$ of $\SU(9)_{\rm TF}$, that decomposes as
\begin{eqnarray}
\hbox{TCb : } 240 &=& \nonumber
(1,1)_0 \oplus (1,2)_{\pm 1/2}  \oplus (1,3)_0 \oplus (8,1)_{2 \times 0,-1} \oplus (10,1)_0 \oplus (8,2)_{1/2, 2 \times (-1/2)} \notag \\
&&\oplus (10,2)_{\pm 1/2} \oplus 2(8,3)_0 \oplus (10,3)_0 \oplus (8,4)_{1/2}
\end{eqnarray}
under the SM gauge group
$\SU(3)_c \otimes \SU(2)_L \otimes {\rm U}(1)_Y$. 
The DM candidate is the neutral singlet $QQ\tilde D$, which can be the lightest TCb.

\subsubsection*{l) SU($N$)$_{\rm TC}$ model $\Q = Q\oplus \tilde D\oplus \tilde U$}
This extension of the previous model allows for two Yukawa couplings, 
$Q H \tilde D$ and $Q H \tilde U$, so that there are no stable TC$\pi$.
This model has $\NF=12$ and is allowed only for $N=3$, where $\Delta b_3 = 8$. 
It predicts an extended set of unstable TC$\pi$,
that fills a $143$ of $\SU(12)_{\rm TF}$:
\beq \hbox{TC$\pi$ : } 143=
\hbox{TC$\pi$}_{Q \oplus \tilde D} \oplus (1,1)_{0,\pm 1} \oplus (1,2)_{\pm 1/2} \oplus (8,1)_{0,\pm 1} \oplus (8,2)_{\pm 1/2} \eeq
under $\SU(3)_c \otimes \SU(2)_L \otimes {\rm U}(1)_Y$.
The model contains two  TCb DM candidates: $QQ \tilde D$ and $\tilde D \tilde D \tilde U$.
The lighter TCb fill a $572$ of $\SU(12)_{TF}$, that decomposes as
\begin{eqnarray}
\hbox{TCb : } 572 &=& \hbox{TCb}_{Q \oplus \tilde D} \oplus (1,1)_{0,2 \times 1} \oplus (1,2)_{2 \times 1/2,3/2} \oplus (1,3)_1 \oplus (8,1)_{2 \times 0, 4 \times 1,2} \oplus (10,1)_{0, 2 \times 1} \notag \\
&& \oplus  (8,2)_{4 \times 1/2, 2 \times 3/2} \oplus (10,2)_{2 \times 1/2,3/2} \oplus 2 \times (8,3)_1 \oplus (10,3)_1
\end{eqnarray}
under $\SU(3)_c \otimes \SU(2)_L \otimes {\rm U}(1)_Y$. 
The TCb DM candidates are those of the $\Q = Q\oplus \tilde D$ model plus a singlet made of $\tilde D \tilde D \tilde U$.

%

\bigskip

Notice that colored techni-quarks never provide golden-class models for $N\ge 4$.
For example the model $\Q=G$ leads, for $N=4$, to an acceptable TCb DM candidate; but $g_3$ develops a sub-Planckian Landau pole.\footnote{Detailed group-theoretical computations show that the simplest model $\Q=G$ leads, for $N=3$ only to coloured lighter TCb;
this can be cured by adding extra techni-quarks (e.g. $\Q = G\oplus N$) but their addition lead to stable coloured TC$\pi$ or TCb.  Furthermore $g_3$ develops a Landau pole below the Planck scale.
So these are not golden-class models. } 
Landau poles also exclude the model $\Q = Q\oplus \tilde U \oplus \tilde D$
(two Yukawa couplings allowed, no stable TC$\pi$) that for $N=6$ provides a TCb DM candidate, $QQ\tilde D\tilde D\tilde D\tilde U$.

\section{SO($N$)$_{\rm TC}$ Composite Dark Matter models}\label{SO}

In this section we consider models based on $\SO(N)$ techni-color interactions with techni-quarks in the vector representation of  $\SO(N)$\footnote{We do not consider Sp$(N)$ techni-color interactions, since there are no stable techno-baryons: the anti-symmetric combination of $N$ techni-quarks decays into $N$ techni-mesons. We also ignore models with chiral representations of the gauge group, which lead to more complicated patterns of symmetry breaking that are not under good theoretical control.  Our results partly hold also for fermions in more general real representations,  but TCb may have different properties~\cite{bolognesi}. }. The techni-quark content is restricted by demanding that $g_{Y,2,3}$ do not develop sub-Planckian Landau poles,
and that  $\SO(N)_{\rm TC}$ is asymptotically free.   Normalizing the generators in the fundamental as $\Tr(T^a T^b) = \delta^{ab}$, 
the $\SO(N)_{\rm TC}$ $\beta$-function coefficient reads
\begin{equation} 
b_{\rm TC} =  - \frac{11}{3}(N-2) + \frac23 \NF <0\qquad  \hbox{so that $\displaystyle\NF <  \frac{11}{2}(N-2)$}.
\end{equation}

Considering again techni-quarks in fragments of the simplest SU(5) representations in table~\ref{tab:unificaxion}, vectorial techni-quarks $\Q$ are defined as:
\beq \Q \equiv  \left\{\begin{array}{ll}
C_N \oplus \bar C_N & \hbox{for complex SM representations $C \in  \{E,L,D,U,Q,S,T,X\}$}\\
R_N & \hbox{for real SM representations $R\in \{N,V,G\}$}
\end{array}\right.  \ .
\eeq
The dynamics of the theory is as follows.  In the limit of negligible techni-quarks masses, the anomaly free global symmetry is $\SU(\NF)\otimes \Z_{(3+(-1)^N)\NF}$
which is spontaneously broken to $\SO(\NF)\otimes \Z_2$  by the condensates 
\begin{equation}
\langle C_N \bar C_N \rangle=2 \langle R_N  R_N \rangle \sim 4\pi \Lambda_{\rm TC}^3  \,.
\label{SONcondensate}
\end{equation}
The spontaneous breaking produces  $\NF(\NF+1)/2-1$ pseudo-Goldstone bosons that transform in the 
two-index symmetric representation of the unbroken $\SO(\NF)$ group.
The condensate preserves the accidental $\U(1)$ symmetry rotating $C_N$ and $\bar C_{{N}}$  with opposite phases, that generalises the species symmetry defined for $\SU(N)_{\rm TC}$ theories.  


The important novelty of this class of models is that the technicolor representation is real. 
This has various consequences: TC$\pi$ are $\Q\Q$ states and there is no distinction between TCb and anti-TCb.
Moreover $N,V,G$ techni-quarks lie in real representations under both $G_{\rm SM}$ and $\SO(N)_{\rm TC}$ and can have Majorana masses
that do not arise in $\SU(N)_{\rm TC}$ models. 

\subsection{Techni-pions and techni-baryons of SO($N$)$_{\rm TC}$} \label{sec:SON-pions-baryons}
There are important differences with respect to  $\SU(N)_{\rm TC}$ models.

Techni-pions are now $\Q\Q$ states, such that, if species number is conserved,
TC$\pi$ made of $C_N C_N$ are stable because they have species number 2.
Furthermore they have quantum numbers under the SM gauge group not compatible with DM phenomenology.
Real techni-quarks $R_N$ instead do not produce stable TC$\pi$ since the techni-quark condensate and masses break their species number.

The presence of at least one techni-quark in a real representation is a necessary ingredient to build viable models 
without unwanted stable TC$\pi$. In fact, Yukawa couplings of the form 
$H R_N C_N$  can break the unwanted species symmetries allowing all TC$\pi$ to decay.
The allowed Yukawa interactions with the Higgs are (analogously to eq.\eq{SUyuk}):
\beq 
HL  ({E}\hbox{ or }{T}\hbox{ or }N\hbox{ or }V),\qquad
HQ  ({D}\hbox{ or }{U}),\qquad
HD  {X}.\qquad
\eeq
$G$-parity can still be defined as in $\SU(N)_{\rm TC}$ theories.
However, with our choice of representations,  $G$-parity is only conserved by the SM multiplet $V$  that in 
$\SO(N)_{\rm TC}$ theories only gives rise to (unstable)  $G$-even TC$\pi$.

\bigskip

Techni-baryons (TCb)  are, as in $\SU(N)_{\rm TC}$ theories, antisymmetric combinations of $N$ techni-quarks.
Techni-baryon number is not conserved, such that TCb cannot have an asymmetry,
two TCb can annihilate and TCb can now be real particles, e.g.\ Majorana fermions. 
The lightest TCb is stable and can be a DM candidate. For $N$ odd stability simply follows from the accidental \mbox{$\Q\to -\Q$} symmetry.
For generic $N$ stability follows  because the $\SO(N)$ gauge theory actually has an accidental ${\rm O}(N)$ symmetry; 
the quotient $\Z_2 = {\rm O}(N)/\SO(N)$ (that distinguishes  orthogonal matrices
according to the sign of their determinant) acts as a global symmetry group.
All TCb  built with the $N$-index anti-symmetric tensor are odd under this $\Z_2$ symmetry,
and the lightest odd state is stable. 

\medskip

Since the same anti-symmetric tensor with $N$-indices is invariant under both $\SU(N)_{\rm TC}$ and
$\SO(N)_{\rm TC}$, the TCb following from a given set of techni-quarks are the same.
They must however be decomposed under different techni-flavor groups conserved by technicolor
interactions: $\SU(\NF)$ for $\SU(N)_{\rm TC}$, and $\SO(\NF)$ for $\SO(N)_{\rm TC}$.
Since $\SO(\NF)\subset \SU(\NF)$, one can start from the  TCb of $\SU(\NF)$ and split them into $\SO(\NF)$ multiplets. The group-theoretic decomposition rules that connect the TCb representations of $\SU(\NF)$ and $\SO(\NF)$ are the following:\footnote{The information contained in these $\SO(\NF)$ Young diagrams is redundant for small $\NF$. Only diagrams with as many rows as the rank of the corresponding $\SO(\NF)$ group are independent. The rank of $\SO(\NF)$ is $\NF$/2 for $\NF$ even and $(\NF-1)$/2 for $\NF$ odd.}
\begin{eqnarray}
N=3 &:& \qquad {\tiny \Bigl(\Yvcentermath1 \yng(2,1)\Bigr)_{\SU(\NF)} \ =\  \Bigl( \yng(2,1) \oplus\yng(1)\Bigr)_{\SO(\NF)} } 
\notag \\
N=4 &:& \qquad  {\tiny \Bigl(\Yvcentermath1 \yng(2,2)\Bigr)_{\SU(\NF)}\ = \  \Bigl(\yng(2,2)\oplus \yng(2)\oplus {1}\Bigr)_{\SO(\NF)} } 
\notag \\
N=5 &:& \qquad {\tiny \Bigl(\Yvcentermath1 \yng(3,2)\Bigr)_{\SU(\NF)}\ = \  \Bigl(\yng(3,2)\oplus\yng(2,1)\oplus \yng(3)\oplus\yng(1)\Bigr)_{\SO(\NF)} } \,.
\label{SOrules}
\end{eqnarray}
This leads to a novel physical phenomenon: $\SO(N)_{\rm TC}$ gives different masses
 to the TCb multiplets that were degenerate in $\SU(N)_{\rm TC}$ models.
For example, in ordinary QCD, if color $\SU(3)$ were replaced by $\SO(3)$ (with 3 quarks in its real fundamental representation),
the `eightfold way' would split into `threefold way' and `pentafold' way:
\beq 8={\tiny \Bigl(\Yvcentermath1 \yng(2,1)\Bigr)_{\SU(3)} \ =\  \Bigl( \yng(2,1) \oplus \yng(1)\Bigr)_{\SO(3)} }=5\oplus 3 \,, \eeq
with a similar decomposition for the heavier  decuplet of spin-3/2 baryons:
\beq
10= {\tiny \Bigl(\yng(3)\Bigr)_{\SU(3)} = \Bigl(\yng(3)\oplus \yng(1)\Bigr)_{\SO(3)}}= 7\oplus 3 \,. \eeq

Unfortunately, QCD gives us no guidance in understanding a crucial question for
composite DM phenomenology:  {\em which $\SO(\NF)$ multiplet contains the lighter TCb,
given that more representations have the same spin?}

Given that composite spin-1 resonances behave as gauge vectors of the techni-flavor symmetries,
and that gauging of global symmetries likely generates positive contributions to  TCb masses,
a plausible answer is that the lightest TCb multiplet is the one in the smallest representation
of $\SO(\NF)$ among those with lowest spin. We will make this assumption in what follows
(dedicated lattice simulations could check this, present results do not allow to settle the issue~\cite{DeGrand:2015lna}).
This means that for $N$ odd the lightest  TCb will be in the vectorial representation of $\SO(\NF)$ (denoted by ${\tiny \Yvcentermath1 \yng(1)}$) 
with the same quantum numbers as techni-quarks $\Psi$ itself, while for even $N$ it will be a singlet of $\SO(\NF)$.

\medskip

Even within the assumption above, if techni-quark masses are comparable to $\Lambda_{\rm TC}$, it becomes possible  that the lightest TCb  belongs to a higher $\SO(\NF)$ representation.
For completeness, we therefore also specify the SM decomposition of the higher $\SO(\NF)$ representations appearing in eq.~(\ref{SOrules}). Notice that for $N=4$, the ${\tiny \Yvcentermath1 \yng(2)}$ representation coincides with the representation of the TC$\pi$, so we only need to specify the ${\tiny \Yvcentermath1 \yng(2,2)}$ representation. Analogously, for $N=5$ we only need to decompose ${\tiny \Yvcentermath1 \yng(3,2)}$ and ${\tiny \Yvcentermath1 \yng(3)}$.


\medskip

Finally,  the members of the lightest TCb $\SO(\NF)$ multiplet are further split by SM gauge interactions and the lightest TCb is the one with the smallest SM charge.

%
%

\begin{table}[t]
\footnotesize
$$\begin{array}{ccc|cc|c}
\multicolumn{1}{c}{\hbox{$\SO(N)$ techni-color.}} & \hbox{Yukawa}&\hbox{Allowed}
&\hbox{Techni-}& \multicolumn{1}{c|}{\hbox{Techni-}} &   \\  
\multicolumn{1}{c}{\hbox{Techni-quarks}}&\hbox{couplings}&\hbox{$N$} &
 \hbox{pions} & \hbox{baryons} &\hbox{under} \\
 \hline
%
\rowcolor[cmyk]{0,0,0.2,0.1}  {\NF=3} &&&  5 &3,1,...\hbox{ for $N=3,4,...$}&  \hbox{$\SO(3)_{\rm TF}$}  \\ 
\rowcolor[cmyk]{0,0,0.2,0} \Q =  V  &0&3,4,..,7 & \hbox{unstable} & V^N= 3,1,... & \hbox{$\SU(2)_L$}\\   \hline
\rowcolor[cmyk]{0,0,0.2,0.1}  {\NF=4} &&&  9 &4,1,...&  \hbox{$\SO(4)_{\rm TF}$}  \\ 
\rowcolor[cmyk]{0,0,0.2,0}  \Q =  N \oplus V  &0&3,4,..,7 & \hbox{3} &VVN=1,V(VV+NN)=3, & \hbox{$\SU(2)_L$}\\ 
\rowcolor[cmyk]{0,0,0.2,0}   & &  & & VV(VV+NN)=1,... & \hbox{$\SU(2)_L$}\\
  \hline
\rowcolor[cmyk]{0,0,0.2,0.1}  \NF=5 & && 14 &5,1...&  \hbox{$\SO(5)_{\rm TF}$}  \\ 
\rowcolor[cmyk]{0,0,0.2,0} \Q =  L\oplus N  &1&3,4,..,14& \hbox{unstable} & L\bar L N =1, & \hbox{$\SU(2)_L$}\\
\rowcolor[cmyk]{0,0,0.2,0}  & & & & L \bar L(L \bar L+NN)=1,... & \hbox{$\SU(2)_L$}\\   \hline
%
%
\rowcolor[cmyk]{0,0,0.2,0.1}   \NF=7 & & & 27  &1,... &  \hbox{$\SO(7)_{\rm TF}$}  \\ 
\rowcolor[cmyk]{0,0,0.2,0}  \Q =  L\oplus V &1&4& \hbox{unstable} & (L\bar L + VV)^2=1 & \hbox{$\SU(2)_L$}\\   \hline
\rowcolor[cmyk]{0,0,0.2,0}  \Q =  L\oplus E\oplus N &2&4,5& \hbox{unstable} & (E\bar E+L\bar L)^2+NN(L\bar{L}+E\bar{E})=1& \hbox{$\SU(2)_L$}\\   \hline
\rowcolor[cmyk]{0,0,0.2,0.1}   \NF=8 & & & 35  &1&  \hbox{$\SO(8)_{\rm TF}$}  \\ 
\rowcolor[cmyk]{0,0.2,0,0} \Q =  G &0& 4&\hbox{unstable} &  GGGG=1& \hbox{$\SU(2)_L$}\\   \hline
\rowcolor[cmyk]{0,0,0.2,0}  \Q =  L\oplus N\oplus V &2&4& \hbox{unstable} & (L\bar L+V V)^2+NN(L\bar{L}+ VV)=1 & \hbox{$\SU(2)_L$}\\   \hline
\rowcolor[cmyk]{0,0,0.2,0.1}   \NF=9 & & & 44  &1&  \hbox{$\SO(9)_{\rm TF}$}  \\ 
\rowcolor[cmyk]{0,0,0.2,0}  \Q =  L\oplus E\oplus V &2&4& \hbox{unstable} & (E\bar E+L\bar L+VV)^2=1 & \hbox{$\SU(2)_L$}\\   \hline
\rowcolor[cmyk]{0,0,0.2,0.1}   \NF=10 & & & 54  &1&  \hbox{$\SO(10)_{\rm TF}$}  \\ 
\rowcolor[cmyk]{0,0,0.2,0}  \Q =  L\oplus E\oplus V\oplus N &3& 4&\hbox{unstable} & \hspace{-1ex} \hbox{as $L\oplus E\oplus V $}+  NN(L\bar{L}+ E\bar{E}+VV)=1 \hspace{-2ex}& \hbox{$\SU(2)_L$}\\   \hline
\end{array}
$$
\caption{\em Golden-class models with $\SO(N)_{\rm TC}$ techni-color. Notations are as in table~\ref{dmsummaryfinal}.  In various models the DM candidate is a linear combination of states.
\label{dmsummaryfinalSO}}
\end{table}

\subsection{SO($N$)$_{\rm TC}$ golden-class  models}\label{sec:SON-models}

As discussed above, avoiding unwanted stable TC$\pi$ implies that the model must
contain at least one real $V$, $N$, $G$ state with Majorana mass.
This leads to real DM states,
with important consequences for DM phenomenology discussed in section~\ref{MajDM}.
With the assumption that the lightest TCb multiplet is the one in the smallest representation of $\SO(\NF)$ among those with lowest spin, table~\ref{dmsummaryfinalSO} lists the golden-class models discussed below.
These are the models that give a DM candidate without  unwanted stable particles.
In appendix \ref{sec:silvermodels} we will present the silver-class models that need extra assumptions  to break  accidental symmetries in order to avoid unwanted stable states.


\subsubsection*{a) SO($N$)$_{\rm TC}$ model $\Q=V$}
This model has $\NF=3$;
TC$\pi$ are unstable, as they lie in the $G$-even representation $5_0$ under $\SU(2)_L\otimes\U(1)_Y$. 
Landau poles are avoided for $N\le 7$ and TCb DM candidates $V^N$ exist for any $N$. 
For $N=3$, the lightest DM candidate has spin $1/2$ and lies in the $3_0$ representation while the heavier TCb lie in the ${\tiny \Yvcentermath1 \yng(2,1)} = 5_0$ multiplet. 
For $N=4$ the TCb DM candidate is a scalar singlet. Also, we have heavier TCb in the ${\tiny \Yvcentermath1  \yng(2)}= 5_0$ representation of $\SU(2)_L$, while the representation ${\tiny \Yvcentermath1  \yng(2,2)}$ is absent.
Finally, for $N=5$, the lightest DM candidate is a $3_0$ multiplet with spin 1/2, the heavier TCb are  a ${\tiny \Yvcentermath1 \yng(2,1)}=5_0$ multiplet and a ${\tiny \Yvcentermath1 \yng(3)} = 7_0$ multiplet while the {\tiny \Yvcentermath1 \yng(3,2)} representation is absent.




\subsubsection*{b) SO($N$)$_{\rm TC}$ model $\Q=N \oplus V$}
This extension of the previous model has $\NF=4$ and it is allowed up to $N =7$. 
The model gives an extended list of TC$\pi$
\beq
\hbox{TC$\pi$ : } 9 = 1_0 \oplus 3_0 \oplus 5_0 \qquad \hbox{under $\SU(2)_L \otimes \U(1)_Y$} .
\eeq
The $3_0$ is stable because of species number, giving a  TC$\pi$ DM candidate. 
For $N=3$ the lightest TCb DM candidate lives in the 4-dimensional ${\tiny \Yvcentermath1 \yng(1)}$  representation of $\SO(4)_{\rm TF}$ that is composed by a singlet $NVV$ and a triplet made by a linear combination of $VNN$ and $VVV$.
For $N=4$ the TCb DM candidate is a singlet linear combination of $VVVV$, $VVNN$.
The remaining heavier TCb for $N=3$ are
\beq
\hbox{TCb : } {\tiny \Yvcentermath1 \yng(2,1)  }= 16 =2 \times 3_0 \oplus 2 \times 5_0 \quad \hbox{under $\SU(2)_L \otimes \U(1)_Y$} \,.
\eeq
As explained before, for $N=4$ it is enough to specify the following decomposition
\beq
\hbox{TCb : } {\tiny\Yvcentermath1 \yng(2,2)  }=10 = 2 \times 5_0 \quad \hbox{under $\SU(2)_L \otimes \U(1)_Y$} \,,
\eeq
to describe all possible TCb.

\subsubsection*{c) SO($N$)$_{\rm TC}$ model $\Q=G$}
For $N=4$ this model with $\NF=8$ avoids a sub-Planckian Landau pole for $g_3$ and, at the same time,
techni-color is asymptotically free, $b_{\rm TC}=-2$. 
The model leads to the following colored TC$\pi$, that undergo anomalous decays to gluons:
\beq 
\hbox{TC$\pi$ : } 35=8_0 \oplus 27_0\qquad\hbox{under $\SU(3)_c\otimes\U(1)_Y$} \ .
\eeq
The TCb DM candidate is the SM singlet $GGGG$ and the remaining 
heavier TCb are:
\begin{eqnarray}
{\rm TCb : } 
&&{\tiny\Yvcentermath1 \yng(2,2)  }=300 = 1_0 \oplus 8_0 \oplus 3 \times 27_0  \oplus 64_0 \oplus \Bigl(10_0 \oplus 28_0 \oplus 35_0 \oplus {\rm h.c.}\Bigr)
\end{eqnarray}
under $\SU(3)_c\otimes\U(1)_Y$, plus a set of TCb living in the same representations as the TC$\pi$ above.

\subsubsection*{d) SO($N$)$_{\rm TC}$ model $\Q=L\oplus N$}
This model with $\NF=5$ allows for a Yukawa coupling that involve the neutral state $N$, such that all TC$\pi$ decay.
They fill a 14 of $\SO(5)_{\rm TF}$ that decomposes as
\beq \hbox{TC$\pi$ : } 14 = 3_{\pm1,0}\oplus 2_{\pm 1/2}\oplus 1_0 \qquad\hbox{under $\SU(2)_L\otimes\U(1)_Y$}. \eeq
This model exists for $3\le N \le 14$. 
Let us consider $N=\{3,4,5\}$, for which the lightest TCb are all SM singlets. For example, for $N=3,4$ they are $N L\bar L$ and $L\bar L(L\bar L+NN)$ respectively. To specify the complete set of TCb, we need the following decompositions 
\beq \label{SO:LN}
\hbox{TCb : }\left\{\begin{array}{ll}
{\tiny\Yvcentermath1 \yng(2,1)  }=35=1_{0,\pm 1}\oplus  2_{\pm 1/2,\pm 1/2,\pm 3/2} \oplus   3_{2 \times 0,\pm1}\oplus 4_{\pm 1/2}
&\hbox{for $N=3$}
 \\
{\tiny\Yvcentermath1 \yng(2,2)  }=35=1_{0,\pm 1, \pm 2}  \oplus 2_{\pm 1/2, \pm 3/2} \oplus 3_{0, \pm 1} \oplus 4_{\pm 1/2} \oplus 5_0&\hbox{for $N=4$} \\
{\tiny\Yvcentermath1 \yng(3,2)  }\oplus{\tiny \Yvcentermath1 \yng(3)}
= 105\oplus 30 = (6_{\pm 1/2}\oplus 5_{\pm 1,2\times 0}\oplus 4_{\pm 3/2, 2\times (\pm 1/2)}\\
~~\oplus \, 3_{2\times (\pm 1,0),\pm 2}\oplus 2_{2\times (\pm 3/2, \pm 1/2), \pm \frac{5}{2}}\oplus 1_{0,\pm (1, 2)})\oplus\\
~~( 4_{\pm 3/2, \pm 1/2}\oplus 3_{\pm 1,0}\oplus 2_{\pm 1/2}+ 1_0)
&\hbox{for $N=5$}
\end{array}\right.
\eeq 
under $\SU(2)_L\otimes\U(1)_Y$. 
Taking into account the Yukawa couplings, in this model and in the following models
the TCb mix giving real eigenstates which are all good
DM candidates, with a peculiar phenomenology discussed in section~\ref{MajDM}.

In the limit $m_N \gg \Lambda_{\rm TC}$ the $N$ state can be integrated out realizing nicely the silver-class model $\Q=L$
presented in appendix~\ref{sec:silvermodels}.


\subsubsection*{e) SO($N$)$_{\rm TC}$ model  $L\oplus V$}
This model with $\NF=7$
is similar to the $\Q=L\oplus N$ but with a more complex set of TC$\pi$
\beq \hbox{TC$\pi$ : } 27 = 5_0\oplus 4_{\pm 1/2}\oplus 3_{\pm1,0}\oplus 2_{\pm 1/2}\oplus 1_0 \qquad\hbox{under $\SU(2)_L\otimes\U(1)_Y$}. \eeq
The strong coupling $g_{\rm TC}$ is asymptotically free for $N\ge 4$ and $g_2$ avoids a sub-Planckian Landau pole for $N\le 4$
(with $N=5$ slightly excluded).
For $N=4$ the TCb DM candidate is the SM singlet $(L\bar L+VV)^2$ and the remaining heavier TCb decompose under $\SU(2)_L\otimes\U(1)_Y$  as:
\begin{eqnarray}
\hbox{TCb : } {\tiny\Yvcentermath1 \yng(2,2)  }={168}&=& 1_{3\times 0,\pm 1,\pm 2} \oplus 2_{3 \times (\pm 1/2),\pm 3/2} \oplus 3_{4 \times 0,3 \times (\pm 1)} \oplus 4_{4 \times (\pm 1/2), \pm 3/2} \notag \\
&& \oplus 5_{4 \times 0, \pm 1} \oplus 6_{2 \times 1/2} \oplus 7_{0, \pm 1} \qquad\hbox{under $\SU(2)_L\otimes\U(1)_Y$} \ .
\end{eqnarray}

\subsubsection*{f) SO($N$)$_{\rm TC}$ model $\Q=L\oplus E\oplus N$} 
This model with $\NF=7$ and 2 Yukawa couplings $HLN$ and $HLE$ predicts the following unstable TC$\pi$
\beq \hbox{TC$\pi$ : } 27 = 3_{\pm 1,0}\oplus 2_{\pm 3/2}\oplus 2\times 2_{\pm 1/2}\oplus 1_{\pm2,\pm 1}\oplus 2\times 1_0 \qquad\hbox{under $\SU(2)_L\otimes\U(1)_Y$}. \eeq
The model exists for $N=4,5$.
For $N=4$ the DM candidate is a singlet, then to fully specify the complete set of TCb we need to decompose the multiplet:
\begin{eqnarray}
\hbox{TCb : } {\tiny\Yvcentermath1 \yng(2,2)} =168 &=&  5_0\oplus 4_{\pm 3/2,2\times (\pm 1/2)}\oplus 3_{\pm 3 ,2\times (\pm 2), 5\times (\pm 1), 5\times 0}\nonumber \\
&&\oplus\, 2_{2\times (\pm 5/2), 5\times (\pm 3/2),7\times (\pm 1/2)} \oplus 1_{3\times(\pm 2),4\times (\pm 1), 6\times 0}  \
\end{eqnarray}
under $\SU(2)_L\otimes\U(1)_Y$.

\subsubsection*{g) SO($N$)$_{\rm TC}$ model $\Q=L\oplus V\oplus E$}
This model with $\NF=9$ and 2 Yukawa couplings $HLV$ and $HLE$, gives rise to the set of unstable TC$\pi$:
\beq \hbox{TC$\pi$ : } 44 =5_0\oplus 4_{\pm 1/2} \oplus 3_{2\times (\pm 1), 0}\oplus 2_{\pm 3/2, 2\times (\pm 1/2)} \oplus 1_{\pm 2}\oplus 2\times 1_0 \qquad\hbox{under $\SU(2)_L\otimes\U(1)_Y$}. \eeq
The model exists and gives a singlet TCb DM candidate for $N=4$.
The multiplet of the remaining heavier TCb  is:
\begin{eqnarray}
\hbox{TCb : } {\tiny\Yvcentermath1 \yng(2,2)} &&= 495 = 1_{8 \times 0,4 \times (\pm 1),3 \times (\pm 2)} \oplus 2_{10 \times (\pm 1/2),6 \times (\pm 3/2),2\times (\pm 5/2)} \oplus 3_{11 \times 0,10 \times (\pm 1),3 \times (\pm 2), \pm 3}  \notag \\
&& \oplus 4_{9 \times (\pm 1/2),5 \times (\pm 3/2),\pm 5/2}
\oplus 5_{7 \times 0,4 \times (\pm 1), 2 \times (\pm 2)} \oplus 6_{3 \times (\pm 1/2), \pm 3/2} \oplus 7_{0,\pm 1} 
\end{eqnarray}
under $\SU(2)_L\otimes\U(1)_Y$.

\subsubsection*{h) SO($N$)$_{\rm TC}$ model $\Q=L\oplus V\oplus N$}
This model has $\NF=8$ and 2 Yukawa couplings ($HLV$, $HLN$) are allowed, so that all TC$\pi$ decay:
\beq \hbox{TC$\pi$ : } 35 =5_0\oplus 4_{\pm 1/2} \oplus 3_{\pm 1,0, 0}\oplus 2\times 2_{\pm 1/2} \oplus  2\times 1_0 \qquad\hbox{under $\SU(2)_L\otimes\U(1)_Y$}. \eeq
The model is allowed only for $N=4$ and gives a singlet TCb DM candidate.
The complete set of TCb contains the multiplet:
\begin{eqnarray}
\hbox{TCb : } {\tiny\Yvcentermath1 \yng(2,2)  } &=& 300 = 7_{\pm 1,0}\oplus 6_{3\times (\pm 1/2)} \oplus  5_{2\times (\pm 1), 8\times 0}\oplus  4_{\pm 3/2, 8\times (\pm 1/2)} \oplus\  3_{6\times (\pm 1), 9\times 0} \nonumber \\
&&\oplus  2_{2\times (\pm 3/2), 7\times (\pm 1/2)}  \oplus 1_{\pm 2, 2\times (\pm 1), 6\times 0}  \qquad\hbox{under $\SU(2)_L\otimes\U(1)_Y$}.
\end{eqnarray}

\subsubsection*{i) SO($N$)$_{\rm TC}$ model $\Q=L\oplus E\oplus V\oplus N$} 
This model with $\NF=10$ and 3 Yukawa couplings $HLV$, $HLN$, $HLE$ predicts the following unstable TC$\pi$:
\beq \hbox{TC$\pi$ : } 54 =5_0\oplus 4_{\pm 1/2} \oplus 2\times 3_{\pm 1,0}\oplus 2_{\pm 3/2, 3\times (\pm 1/2)} \oplus 1_{\pm 2, \pm 1}\oplus 3\times 1_0 \qquad\hbox{under $\SU(2)_L\otimes\U(1)_Y$}. \eeq
For $N=4$, the model gives the singlet TCb DM candidates, while the multiplet of the remaining heavier TCb  is:
\begin{eqnarray}
\hbox{TCb : } {\tiny\Yvcentermath1 \yng(2,2)  } &=& 770 =1_{5 \times (\pm 2), 8 \times (\pm 1), 14 \times 0} \oplus 2_{3 \times (\pm 5/2), 11 \times (\pm 3/2), 19 \times (\pm 1/2)} \oplus 3_{\pm 3, 5 \times (\pm 2), 17 \times (\pm 1), 20 \times 0}  \notag \\
&&\oplus 4_{\pm 5/2, 7 \times (\pm 3/2), 15 \times (\pm 1/2)} \oplus 5_{2 \times (\pm 2), 6 \times (\pm 1), 11 \times 0} \oplus 6_{\pm 3/2, 4 \times (\pm 1/2)} \oplus 7_{\pm 1, 0} 
\end{eqnarray}
under $\SU(2)_L\otimes\U(1)_Y$.

\section{Techni-quark masses and the $\theta_{\rm TC}$ angle}\label{spec}
In \cite{us} we considered composite dark matter theories in the limit of massless techni-quarks.
With masses (such that also the CP-violating $\theta_{\rm TC}$ angle becomes physical) the theory has a few more free parameters, 
that significantly affect its phenomenology. 
From a phenomenological point of view, we are mostly interested in checking that a successful TCb DM  candidate is indeed the lightest TCb 
and in computing its interactions. The main new feature relevant for DM direct detection is that  DM TCb fermion  
has magnetic and electric dipoles with moments
\begin{equation}
 \bar\Psi \gamma_{\mu \nu} (\mu_M + i d_E  \gamma_5) \Psi ~   F_{\mu\nu}/2 .
\label{dipoles}
\end{equation}
We estimate
\beq  \mu_M \sim \frac{e}{M_{\rm DM}},\qquad d_E \sim  \frac {e\, \theta_{\rm TC}}{16\pi^2}\frac {\min[m_\Q]}{f^2}
\sim  \frac {e\, \theta_{\rm TC} \min [m_\Q]}{M_{\rm DM}^2} \, .
\eeq
A magnetic moment with order 1 gyro-magnetic ratio is typical of composite states. The smaller electric dipole is generated when CP is violated 
by a non-zero $\theta_{\rm TC}$. For $\theta_{\rm TC}\sim {\cal O}(1)$ EDM could give striking effects in direct detection 
as we will see in section \ref{pheno}. Chromo-dipoles are  generated in models with colored constituents.

\subsection{A QCD-like example}\label{QCDlike}

To illustrate the effects of the $\theta_{\rm TC}$ angle, assumed to be large unlike the QCD $\theta$-angle, 
we work out in detail the silver-class model with $\SU(3)_{\rm TC}$  and $\Q=L\oplus E$ techni-quarks, described in section~\ref{modelLE}.
In this scenario the techni-strong dynamics is identical to QCD with three flavors and therefore we can rescale QCD data to make definite predictions. 
For this choice of quantum numbers  no Yukawa couplings are allowed, such that charged TC$\pi$ are stable at renormalizable level.  
We assume that non-renormalizable operators  break species number symmetry leading to unstable TC$\pi$, and that DM is the singlet neutral TCb.

The TC$\pi$ in the adjoint of $\SU(3)_{\rm TF}$ and the anomalous $\U(1)$ singlet are described by the hermitian matrix
\begin{equation}
\Pi = \left( \begin{array}{ccc}
\pi_3^0/\sqrt{2}+\pi^0_1/\sqrt{6}& \pi_3^+ & \pi_2^- \\
\pi_3^{-} & -\pi_3^0/\sqrt{2}+\pi^0_1/\sqrt{6} & \pi_2^{--} \\
\pi_2^+ & \pi_2^{++} & -2\pi^0_1/\sqrt{6} \end{array} \right) + \frac{\eta'}{\sqrt{3}} 1\hspace{-0.9ex}1_3.
\label{QCDpions}
\end{equation}
This is as in QCD, but with  different charges for the isospin doublets. The effective TC$\pi$ Lagrangian described in \cite{pich} reads
\begin{equation}
\mathscr{L}_{{\rm TC}\pi}\approx\frac{f^2}{4} \bigg\{{\rm Tr}[D_\mu U D^\mu U^\dagger] + 
2 B_0 {\rm Tr}[M (U +U^\dagger)]
-\frac{a}{3} \bigg[ \theta_{\rm TC}- \frac{i}{2}(\ln \hbox{det}\ U- \ln \hbox{det}\ U^\dagger)\bigg]^2 \bigg\} \, ,
\label{anomaly}
\end{equation}
where $U=\langle U \rangle e^{i \sqrt{2} \Pi/f}$ is the TC$\pi$ matrix. The second term in the lagrangian describes the effect of techni-quark masses, where 
 $M={\rm diag}(m_L, m_L, m_E)$, and $B_0$ is the chiral condensate. The last term encodes the effect of the $\theta_{\rm TC}$ angle and $\U(1)$ axial anomaly that gives mass to the techni-$\eta'$,
$m_{\eta'}^2\sim a +{\cal O}(m)$. 

The VEV $\langle U\rangle$  is determined dynamically by minimising the potential.
One can conveniently look for a solution of the form
\begin{equation}
\langle U \rangle=\hbox{diag}\,(e^{-i\phi_L},e^{-i\phi_L},e^{-i\phi_{E}})\,.
\label{Uvev}
\end{equation} 
The extrema of the potential are determined by the Dashen's equations:
\begin{equation}
\chi^2_L \sin\phi_L = \frac{a}{3} (\theta_{\rm TC} -2\phi_L-\phi_E) \,, \qquad \chi_E^2 \sin\phi_E = \frac{a}{3} (\theta_{\rm TC} -2\phi_L-\phi_E) \,,
\label{min-2+1}
\end{equation}
where we defined $\chi_{E,L}^2 \equiv -2 m_{E,L} B_0$. It is easy to check that  $\langle U\rangle \ne 1\hspace{-0.9ex}1$ when $\theta_{\rm TC}\ne 0$ and techni-quark masses are different
from zero. A non-vanishing $\theta_{\rm TC}$ modifies the TC$\pi$ spectrum such that $m_{E,L} \to m_{E,L}  \cos \phi_{E,L} $ in the mass formul\ae{} and  generates CP violating interactions among the TC$\pi$. In the limit $m_\eta'\gg m_\pi$ (corresponding to $\chi_{L,E}^2\ll a$) and neglecting gauge contributions one finds
\begin{equation}
m_{\pi_3}^2= \chi_L^2 \cos\phi_L\,, \quad ~m_{\pi_2}^2 =\frac{\chi_L^2 \cos\phi_L+\chi_E^2 \cos\phi_E}{2}\,, \quad m_{\pi_1}^2=\frac 1 3(\chi_L^2\cos \phi_L+2 \chi_E^2 \cos \phi_E) \,.
\label{pionmass}
\end{equation}
Since $\cos \phi_{E,L} $ can be negative the effect of $\theta_{\rm TC}$ cannot be entirely reabsorbed by redefining the techni-quark masses 
(for example, in real world QCD, the measured pion spectrum is compatible with $\theta=0$ but not with $\theta=\pi$~\cite{wittentheta}). 

\begin{figure}[t]
\begin{center}
\begin{tikzpicture}[line width=1.5 pt, scale=3]
    	\draw[fermion] (0,0)--(0.5,0);
	\draw[fermion] (0.5,0)--(1.5,0);
	\draw[dashed] (0.5,0) arc (180:0:0.5);
	\draw[fill=gray] (0.43,0) arc (180:-180:0.07);
	\draw[fermion] (1.5,0)--(2,0);
    \draw[vector] (1,0.5)--(1,1);
	\node at (-0.05,0.2) {$B_1^0$};
	\node at (2,0.2) {$B_1^0$};
\end{tikzpicture}
\end{center}
\caption{\em Leading logarithmic contribution to the  EDM of the DM candidate $B_1^0$. Internal dashed and continuous lines represent respectively $\pi_2$, $B_2$ states.  The gray dot correspond to the CP violating $\pi B B$ vertex proportional to $b_1$, $b_2$, while the other $\pi B B$ vertex is proportional to the $D$ and $F$ derivative couplings. 
\label{fig:EDM-loop}}
\end{figure}
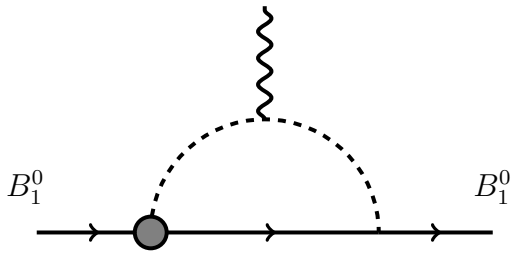

\bigskip

The spectrum of TCb can be computed with similar techniques.
The octet contains
\begin{equation}
B = \left( \begin{array}{ccc}
B_3^0/\sqrt{2}+B_1^0/\sqrt{6} & B_3^+ & B_2^- \\
B_3^- & -B_3^0/\sqrt{2}+B_1^0/\sqrt{6} & B_2^{--} \\
B_{2'}^+ & B_{2'}^{++} & -2 B_1^0/\sqrt{6} \end{array} \right) \,.
\label{QCDbaryons}
\end{equation}
where $B_2$ and $B_2'$ are the analog of the nucleon and the $\Xi$ doublet respectively, $B_3$ of the triplet $\Sigma$ and  $B_1^0$ of the singlet $\Lambda$.
The effective lagrangian for the TCb can be found in \cite{pich}. It contains the following terms relevant to the present discussion:
\begin{eqnarray}
\Lag_{kin} &=& \Tr[\bar B (i\slashed{D} - m_B) B] - 2 \left(b_1\, \Tr[\bar B M_\theta B] + b_2\, \Tr[\bar B B M_\theta]\right) \,, \nonumber \\
\Lag_{BB\Pi,\theta} &=& - \frac{2 \sqrt{2} a}{3 f} \left(\theta_{\rm TC} - 2 \phi_L-\phi_E \right) \left( b_1 {\rm Tr}[\bar B \Pi B] + b_2 {\rm Tr}[\bar B B \Pi] \right) +\dots\,, \nonumber \\
\Lag_{BB\Pi} &=& -\frac{D+F}{\sqrt{2}f}  {\rm Tr}[\bar B \gamma^\mu \gamma_5  (D_\mu \Pi)  B] - \frac{D-F}{\sqrt{2}f}  {\rm Tr}[\bar B \gamma^\mu \gamma_5 B (D_\mu \Pi)]+\dots \,,
\label{baryoneffective}
\end{eqnarray}
where $m_B$ is the common TCb mass generated by the strong interactions 
and $M_\theta$ is the techni-quark mass matrix that depends on $\theta_{\rm TC}$ angle through eq.~(\ref{Uvev})
\begin{equation}
M_\theta = {\rm diag}(\chi_L^2 \cos\phi_L,\, \chi_L^2 \cos\phi_L,\, \chi_E^2 \cos\phi_E).
\end{equation}
The second line of eq.~(\ref{baryoneffective}) describes the CP violating interactions induced by $\theta_{\rm TC}$ relevant for the computation of electric dipoles
and the third line contains derivative interactions with the TC$\pi$. Dots stand for non-linear terms irrelevant for the present discussion.
All the parameters of the effective lagrangian are determined by rescaling QCD data in terms of the dynamical scale,
\begin{eqnarray}
&&\frac {m_\rho} f \sim 8\,,\qquad
{\frac {B_0}{m_\rho}\sim -2}\,,
\qquad
\frac {m_B}{m_\rho}\sim 1.3\,,\qquad \Delta^g m_\pi^2\sim \frac {3 \alpha_2}{4\pi}(J(J+1))\, m_\rho^2\nonumber \\
&&m_B\, b_1 \sim 0.15\,,\qquad m_B\, b_2\sim - 0.3\,,\qquad D\sim0.6 \,,\qquad F\sim 0.4
\label{QCDrescaling}
\end{eqnarray}
where $J$ is the isospin of the TC$\pi$ multiplet.
From the first line of eq.~(\ref{baryoneffective})  the mass splittings between TCb due to techni-quark masses  reads
\begin{eqnarray}
\Delta m_{B_2} &=& 2 (b_1 \chi_L^2 \cos \phi_L+ b_2 \chi_E^2 \cos \phi_E) \,, \quad \Delta m_{B_{2'}}= 2 (b_2 \chi_L^2 \cos \phi_L+ b_1 \chi_E^2\cos \phi_E)\,, \nonumber \\
\Delta m_{B_3} &=& 2 (b_1+ b_2) \chi_L^2 \cos \phi_L \,, \quad  \Delta m_{B_1}=2/3(\chi_L^2 \cos \phi_L+2\chi_E^2 \cos \phi_E)(b_1+b_2) \,.
 \label{baryonoctet}
\end{eqnarray}

The $LLE$ states, corresponding to the triplet $B_3$ and the singlet $B_1$ have zero hypercharge.
Therefore they can be viable DM candidates if they are the lightest TCb.
Using the QCD values of $b_1$ and $b_2$,we find that techni-quark masses always favor $B_2$ or $B_{2'}$ to be the lightest TCb.
The neutral $LLE$ state can be the lightest TCb 
when the mass splitting due to SM gauge interactions is more important than the mass splitting due to techni-quark masses.
This can be realised in the symmetric limit $\chi_L = \chi_E\equiv \chi$ where techni-quark masses respect the techni-flavor symmetry and  the singlet $B_1^0$
(analog of the $\Lambda$) is most likely the lightest TCb.

In the limit  $\chi_L = \chi_E\ll a $  we can solve Dashen's equations analytically. The solution has multiple branches labelled by the integer $n$ \cite{wittenchiral},
\beq
\phi_L = \phi_E- 2  \pi n \simeq \frac{\theta_{\rm TC}-2 \pi n }{3} \, .
\label{Dashen-sol}
\eeq
The solution with minimum energy has a discontinuity at $\theta_{\rm TC}=\pi$ where it jumps from $n=0$ to $n=1$.
This is necessary to restore the periodicity in $\theta_{\rm TC}$.

\subsection{Electric dipole of the DM candidates}

We parameterize dipole moments in terms of gyromagnetic factors  $g_{M,E}$ as
\begin{equation}\label{eq:gfactors}
\mu_M= \frac {e g_M} {2M_{\rm DM}} ,\qquad
d_E= \frac {e g_E} {2M_{\rm DM}}.
\end{equation}
Following \cite{colemanglashow}, to leading order the dipole moments are proportional to the electric charge,
\begin{equation}
\mu(B) = \alpha\, {\rm Tr}[B B^\dagger Q]+ \beta\, {\rm Tr}[B Q B^\dagger]
\label{dipolegrouptheory}
\end{equation}
where $\alpha$ and $\beta$ are properties of the strong dynamics that, for the QCD-like model, 
can be extracted from the measured magnetic moments of baryons in QCD.
What is different in our context is the charge matrix $Q={\rm diag}(0\,,-1\,,1)$. Plugging in the equation above we estimate $g_M^{B_1}\sim  2.8$. 

The same argument applies to the EDMs. To estimate the coefficient we proceed as in \cite{pich} for the computation of the neutron EDM.
The CP violating vertices from the mass terms in eq.~(\ref{baryoneffective}) generate one-loop graphs that contribute to the EDM. 
The dominant contributions are given by the logarithmically divergent diagrams represented in fig.~\ref{fig:EDM-loop}.
Similarly to the computation of the neutron EDM we obtain the estimate,
\begin{equation}
g_E^{B_1} \simeq - \frac{3\chi^2 }{4\pi^2 f^2} M_{\rm DM} \left[b_1(D+F) - b_2 (D-F)\right] \ln\frac {m_{B_2}^2}{m_{\pi_2}^2}\times \sin\frac{\theta_{\rm TC}}{3} \qquad \hbox{for } \theta_{\rm TC} <\pi .
\end{equation}
For $\theta_{\rm TC} \circa{<} 1$ using the numerical values in (\ref{QCDrescaling}) we obtain
\begin{equation}
g_E^{B_1} \simeq -  0.15\, \frac  {m_{\pi_2}^2}{f^2} \log \frac {m_B^2}{m_\pi^2} \times \theta_{\rm TC}.
\end{equation}
The discussion above can be easily generalised to other models. For example the model $\Q=V$ for $N=3$ has again the same dynamics as QCD. 
From eq.~(\ref{dipolegrouptheory})  one can  see that the magnetic and electric dipole moments of the TCb dark matter candidate (the neutral component of an isospin triplet) are zero. 

\bigskip

For different $N$ and $\NF$ the relevant dynamics can be parametrized 
in terms of few unknown parameters that could in principle be extracted from  lattice simulations.
For TC$\pi$ the discussion is identical to eq.~(\ref{anomaly}) with a number of Dashen's equation equal to the number of SM representations of the model.
TCb are in general described by a tensor of $\SU(\NF)$ $B_{i_1\,i_2\dots i_{N}}$ with the symmetry of Young tableaux as in (\ref{youngtableauxlight}).
Their effective lagrangian is constructed writing all possible techni-flavor invariant combinations of the techni-baryon fields $B$ and $\bar{B}$ 
with the techni-quark mass matrix $M$ transforming in the adjoint representation of $\SU(\NF)$.  

For $N$ odd there are two non-trivial invariants: 
\beq
{\rm Tr}[\bar{B} M B] \ ,\ {\rm Tr}[\bar{B} B M] \,.
\eeq 
Since the TC$\pi$ 
are in the adjoint representation,  other two invariants can be written with derivative interactions that do not break 
the global symmetries.

For $N$ even, a single invariant can be written down: 
group theory  uniquely fixes the mass splitting among TCb up to its overall coefficient. For example,
in the model with $N=\NF=4$ we predict equal mass differences between the TCb. 

\section{Phenomenology of Composite Dark Matter}
\label{pheno}

We here briefly outline the phenomenology of the scenarios with TCb dark matter\footnote{If TC$\pi$ are stable due to accidental 
symmetries their mass should not exceed few TeV not to overclose the universe. The TC$\pi$ DM in this case likely dominates 
and behaves as the minimal dark matter candidates studied in \cite{MDM}.}.

This crucially depends on the TCb mass.  Cosmology singles out two special values:
\beq \label{eq:MDMs}
M_{\rm DM}\approx \left\{ \begin{array}{ll}
100\TeV & \hbox{if DM is a thermal relic,}\\
3\TeV & \hbox{if DM is a complex state with a TCb asymmetry \cite{Sannino}.}
\end{array}\right.\eeq
In the first case, the cosmological relic abundance is  determined by the non-relativistic
annihilation cross-section  of TCb, that annihilate into TC$\pi$ through strong interactions and to SM states through gauge interactions. We can neglect the second sub-dominant effect.
Rescaling the measured $p\bar p$ annihilation cross-section one finds \cite{us}
that the thermal DM abundance is reproduced for $M_{\rm DM}\sim 200\TeV$.

\subsection{Direct detection of complex Dark Matter}\label{DiracDM}

In various models, the DM candidate is a complex state with $Y=0$ in the
triplet or quintuplet representation of $\SU(2)_L$. Its weak interactions lead to a direct-detection cross section characteristic of
Minimal Dark Matter, which is too small to be observed in the present context where the DM mass is around 100 TeV,
if DM is a thermal relic.  Moreover in various models DM is a SM singlet, such that even this cross section is absent. 

The main hope for direct detection of thermal TCb DM relies on the fact that composite DM made of charged constituents 
can have special interactions with the photon,
leading to significant rates of low-energy scatterings. Scalar DM $S$ can only have the dimension 6 interaction $(S^* i \partial_\mu S) \partial_\nu F_{\mu\nu}$ or higher, 
which does not lead to interesting rates.  Fermionic DM  $\Psi$ instead can have dipole interactions as in eq.~(\ref{dipoles})
leading  to the following cross section for direct detection~\cite{dip,cirelli}:
\beq \frac{d\sigma}{dE_R} \approx \frac{ e^2 Z^2}{4\pi  E_R}\left(\mu_M^2 + \frac{d_E^2}{v^2}\right) 
\eeq
where $v$ is the relative DM/nucleus velocity and $E_R$ is the nucleus recoil energy. 
For simplicity, we here assumed a nucleus $\cal N$ with $A,Z\gg1$, 
mass $M_{\cal N} \approx A m_N$,
a recoil energy $E_R \ll M_{\cal N} v^2$,
and approximated nuclear form factors with their unit value that holds at small enough $E_R$.
In the same approximation, this cross section can be compared to the standard approximation 
used in searches for spin-independent DM interactions:
\beq \frac{d\sigma}{dE_R} =\frac{ M_{\cal N}\sigma_{\rm SI} A^2}{2\mu^2 v^2},\qquad
\mu= \frac{M_{\cal N} M_{\rm DM}}{M_{\cal N}+ M_{\rm DM}}.
\eeq
We see that the dipole cross sections has a characteristic testable enhancement at low recoil-energy $E_R$,
arising because the DM/matter scattering is mediated by the massless photon.
Furthermore, the magnetic-dipole cross section has a characteristic suppression at small $v > v_{\rm min}=\sqrt{M_{\cal N} E_R/2\mu^2}$, 
which could be tested relying on the seasonal variation in the average $v^2$.

\medskip

We parameterize the dipole moments in terms of their gyro-magnetic and gyro-electric constant $g_M$ and $g_E$ as in eq.\eq{gfactors}.
Composite DM generically predicts an order one gyromagnetic factor $g_M$, and a possibly sizeable
gyro-electric factor $g_E  \sim  \theta_{\rm TC} {\rm Min}[m_{\Q}]/M_{\rm DM}$ as discussed in section \ref{spec}.

\begin{figure}[t]
$$\includegraphics[width=0.45\textwidth]{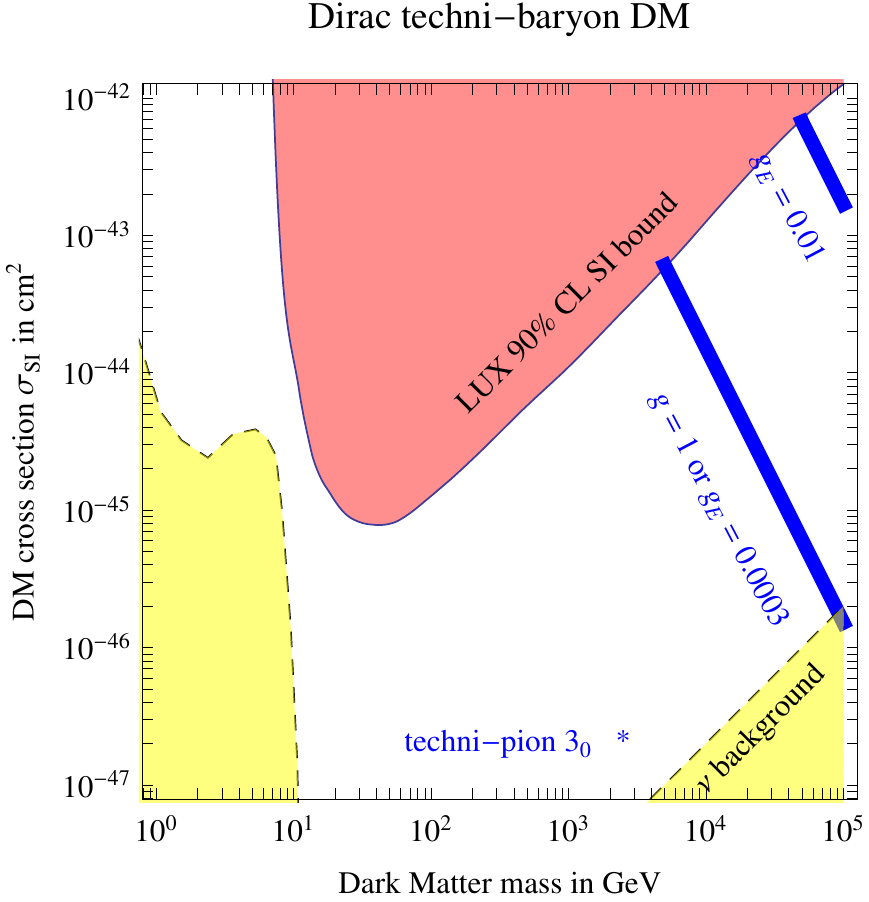}\qquad \includegraphics[width=0.45\textwidth]{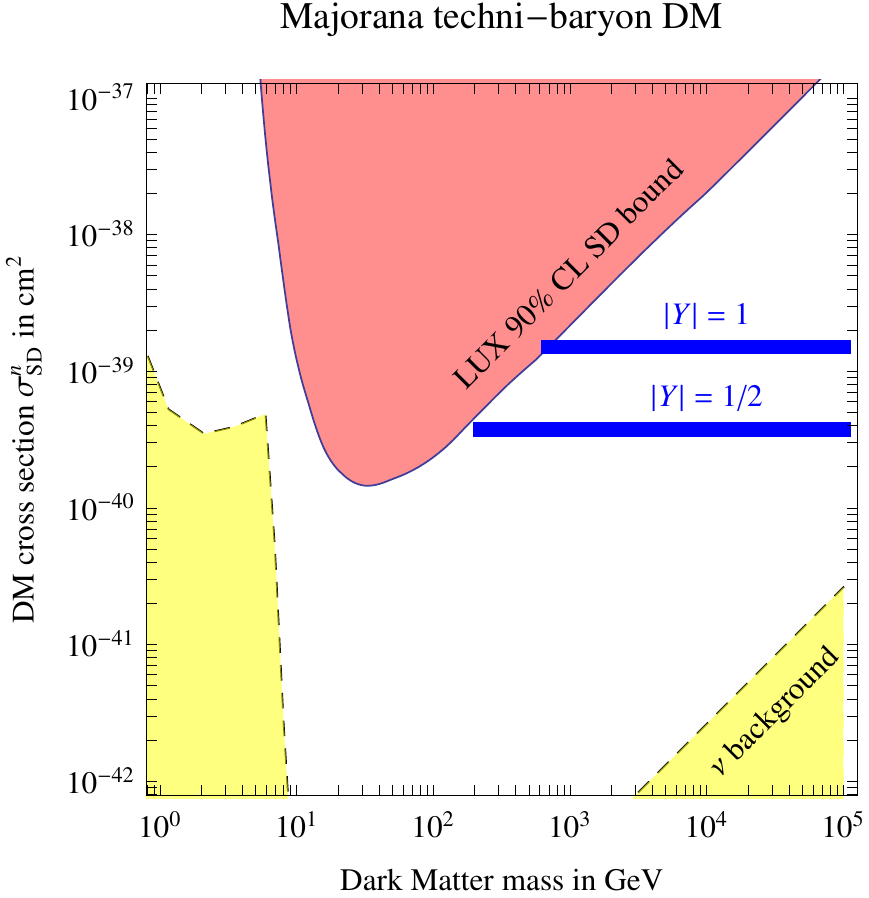}$$
\caption{\em Predictions for direct detection of Dark Matter.  Left: Dirac TCb predicted by $\SU(N)_{\rm TC}$ models to have magnetic 
and/or electric dipole moments, giving spin-independent cross section.
Right: Majorana techni-baryons predicted by $\SO(N)_{\rm TC}$ models to have hypercharge giving spin-dependent cross section.
\label{fig:sigmaDD}}
\end{figure}

This means that for $M_{\rm DM}\approx 100\TeV$ and $g_M\sim 1$ the magnetic effect is 3 orders of magnitude
below the experimental limit,
\beq \sigma_{\rm SI}<10^{-44}\cm^2 \frac{M_{\rm DM}}{\TeV}\qquad
\hbox{for $M_{\rm DM}\gg M_{\cal N}$}\eeq
and at the level of the neutrino background, see also \cite{Appelquist:2013ms}. The electric effect is comparable to the present LUX bound
for $g_E \approx 0.01$ and $M_{\rm DM}\approx 100\TeV$, as illustrated in fig.~\ref{fig:sigmaDD}a.

In some models DM has chromo-dipoles, that lead to a similar scattering rate
with $e^4 Z^2/E_R$ replaced by $g_3^4   /\Lambda_{\rm QCD}$
times a nuclear form factor, which is strongly suppressed at energies below $\Lambda_{\rm QCD}$.
Thereby chromo-dipoles do not compete with electric dipoles.

Some composite DM models predict that DM is a TCb with higher-spin.
Spin 1 DM can have characteristic spin-dependent interactions which are, however, suppressed by the transferred momentum~\cite{Krauss}. 
More interestingly, a composite spin-1 TCb $B_\mu$ could have a dimension-4 interaction
$B_\mu B^*_\nu F^{\mu\nu}$ with a photon. 
Even when the lighter TCb is mostly composed of neutral SM singlets $N$, 
it also contains a small component of charged heavier techni-quarks
with a momentum asymmetry (an effect analogous to the strange momentum asymmetry in nucleons~\cite{NuTeV}).

\subsection{Direct detection of real Dark Matter}\label{MajDM}
Techni-baryon DM in $\SO(N)_{\rm TC}$ gauge theories has novel interesting features compared to $\SU(N)_{\rm TC}$ models: 
there is no  techni-baryon number conservation, so DM is a real state with no techni-baryon asymmetry. In most golden-class models, the techni-quarks have Yukawa couplings to the Higgs.
As a consequence the DM candidates TCb with $Y=0$ mix with  TCb with $Y\neq 0$ after electro-weak symmetry breaking.
The resulting lightest TCb is a Majorana fermion for $N$ odd, a real scalar for $N$ even. 
To illustrate this point, let us consider for example the $L\oplus N$ model with 3 techni-colors. The multiplet of lighter TCb 
in eq.~(\ref{SO:LN}) contains a Majorana singlet $1_0$ and a Dirac weak doublet $2_{\pm 1/2}$.
In view of the Yukawa couplings among the techni-quarks, the mass matrix for the neutral TCb components has the form 
\begin{equation}
\bordermatrix{& 1_0 & 2_{1/2} & 2_{-1/2}  &\cdots \cr
1_0  & m_{1_0} & y_{L} v & y_{R} v & \cdots\cr
2_{1/2}  & y_{L}^* v & 0 & m_{2_{1/2}}  & \cdots\cr
2_{-1/2}  & y_{R}^* v & m_{2_{1/2}} &0& \cdots\cr
\vdots & \vdots & \vdots &  \vdots & \ddots}
 \label{Majorana}
\end{equation}
where the TCb Higgs couplings $y$  can, in principle, be derived from the Yukawa couplings among techni-quarks. The dots refers to other TCb states
that are expected to be heavier but could still be relevant if they mix significantly.

The mass matrix is analogous to one of the bino and higgsino in supersymmetry.
Furthermore, in our scenario TCb have a common mass $m_B$ generated by strong dynamics and are mildly split by techni-quark masses and gauge interactions: thereby the
spectrum resembles the case known as `well tempered neutralinos' \cite{welltempered}. 
Due to the mixing with Majorana states, the lightest DM TCb is a Majorana fermion.
This significantly changes the phenomenology of direct detection: a Majorana fermion can neither have  vector couplings to the $Z$, 
avoiding the severe constraints  from spin independent cross section, nor dipole moments, removing the signals discussed in section~\ref{DiracDM}.
However, Majorana technibaryon DM $\chi$ can have an axial coupling to the $Z$,
\beq\label{Zcouplings}
-g_{A} Z_\mu  \frac{g_2}{\cos\theta_{\rm W}}  \frac{\bar{\chi}\gamma_\mu  \gamma_5  \chi}{2}\, ,
\eeq
that leads to a spin dependent cross-section with the nuclei. Using the present LUX  bound~\cite{LUX} 
$ \sigma_{\rm SD}^n < 1.7 ~10^{-39} M_{\rm DM}/\TeV$, 
one finds
\beq |g_A| < 1.2 \frac{M_{\rm DM}}{\TeV}.\eeq
The situation is illustrated in fig.~\ref{fig:sigmaDD}b.

This is a significant constraint only if the mixing angle among states of different hyper charge 
is large so that $g_A\sim {\cal O}(1)$. This situation is achieved for
\begin{equation}
\Delta m \equiv  |m_{2_{1/2}} -m_{1_{0}}| \circa{<} y v.
\label{largemix}
\end{equation}
Even assuming negligible techni-quark masses, SM gauge interactions split singlets and doublets by a few per cent:
\begin{equation}
\Delta m \approx   \frac {\alpha_2}{4\pi} m_B\sim 0.03\times  m_B.
\end{equation}
For a TCb mass around 100 TeV the condition (\ref{largemix}) is unlikely to be realised:
in the opposite regime $\Delta m\gg y v$ the lightest TCb has suppressed coupling to the $Z$,
\begin{equation}
g_A\sim \frac{y^2 v^2}{\Delta m^2} \ll 1.
\end{equation} 
Another effect of phenomenological relevance can arise if  $m_{2_{1/2}} \ll m_{1_0} $.
In this case the lighter complex doublet splits into two real states, with a mass difference
$
\Delta m_{2_{1/2}} \approx { y^{2} v^2 }/{ \Delta m } 
$.
The $Z$ gives a tree level coupling between the real mass eigenstates, becoming irrelevant 
for direct DM searches if $\Delta m_{2_{1/2}} \circa{>}100\,{\rm keV}$. A smaller mass difference can be obtained for $y \sim 10^{-3}$ 
and gives rise to inelastic DM phenomenology~\cite{TuckerSmith:2001hy}.

\subsection{Higgs-mediated direct detection of Dark Matter}
In both cases (real and complex DM) many golden-class
composite DM models contain Yukawa couplings to the Higgs in order
to break species number symmetries that would lead to unwanted stable particles.
Such Yukawa couplings give rise to an extra Higgs-mediated contribution to the 
spin-independent cross section for direct DM searches, given by
\beq
\sigma_{\rm SI}=\frac{g_{\rm DM}^2 m_N^4 f_N^2}{2\pi v^2 M_h^4}
\eeq
for DM with any spin. Here $f_N\approx0.3$ is a nuclear form factor, $v\approx 174\GeV$ is the Higgs vev,
and $g_{\rm DM}$ is the dimension-less  coupling of the TCb DM candidate with mass $M_{\rm DM}(h)$ to the higgs, defined as
\beq 
g_{\rm DM} = \frac{\partial M_{\rm DM}}{\partial h}\qquad
\eeq
and roughly given by the Yukawa couplings of the Higgs to techni-quarks.
The size of these Yukawa couplings is unknown.
The LUX bound on $\sigma_{\rm SI}$ implies $g_{\rm DM}< \sqrt{M_{\rm DM}/75\TeV}$.

\begin{table}[t]
$$ \begin{array}{c|c|ccc|cc|cc|c|c}
\hbox{gauge}& \hbox{Techni-quark} & \multicolumn{9}{c}{\hbox{Techni-pion content under $\SU(2)_L\otimes\U(1)_Y$}}\\
\hbox{group}& \hbox{content} & 1_0 & 1_{\pm 1} & 1_{\pm 2} & 2_{\pm 1/2} & 2_{\pm 3/2} & 3_0 & 3_{\pm 1} & 4_{\pm 1/2} & 5_0\\ \hline
\rowcolor[cmyk]{0,0,0.05,0}  {\tiny \SU(N)_{\rm TC}} & V                        &   &   &   &   &   & 1_{\rm stable} &  &   & 1 \\
\rowcolor[cmyk]{0,0,0.05,0}  &N\oplus V                 & 1 &   &   &   &   & 3_{\rm stable} &  &   & 1 \\
\rowcolor[cmyk]{0,0,0.05,0} &N\oplus L                 & 1 &   &   & 1 &   & 1              &  &   & \\
\rowcolor[cmyk]{0,0,0.05,0} &N\oplus L \oplus \tilde E & 2 & 1 &   & 2 &   & 1              &  &   & \\
\rowcolor[cmyk]{0,0,0.05,0} &V\oplus L                 & 1 &   &   & 1 &   & 2              &  & 1 & 1\\
\rowcolor[cmyk]{0,0,0.05,0} &V\oplus L\oplus\tilde E   & 2 &   &   & 2 &   & 2              & 1 & 1 & 1\\
\rowcolor[cmyk]{0,0,0.05,0} &V\oplus L\oplus N         & 2 &   &   & 2 &   & 4              &   & 1 & 1\\
\rowcolor[cmyk]{0,0,0.05,0} &N\oplus L \oplus\tilde{L} & 2 & 1 &   & 2 &   & 2              & 1  &   &\\
\rowcolor[cmyk]{0,0,0.05,0} &N\oplus L \oplus\tilde{L}\oplus\tilde{E}         & 3 & 2 &   & 3 & 1 & 2 & 1 &   &  \\
\rowcolor[cmyk]{0,0,0.05,0} &N\oplus L \oplus\tilde{E}\oplus V                & 3 & 1 &   & 3 &   & 4 & 1 & 1 & 1\\
\rowcolor[cmyk]{0,0,0.05,0} &N\oplus L \oplus\tilde{L}\oplus E\oplus\tilde{E} & 4 & 3 & 1 & 4 & 2 & 2 & 1 &   & \\ \hline
\rowcolor[cmyk]{0.05,0,0,0} {\tiny \SO(N)_{\rm TC}}  &V                         &   &   &   &   &   &   &   &   & 1\\
\rowcolor[cmyk]{0.05,0,0,0} &L\oplus N                 & 1 &   &   & 1 &   & 1 & 1 &   &    \\
\rowcolor[cmyk]{0.05,0,0,0} &N\oplus V                 & 1 &   &   &   &   & 1_{\rm stable} &   &   & 1   \\
\rowcolor[cmyk]{0.05,0,0,0} &L\oplus V                 & 1 &   &   & 1 &   & 1 & 1 & 1 & 1  \\
\rowcolor[cmyk]{0.05,0,0,0} &L\oplus N\oplus E         & 2 & 1 & 1 & 2 & 1 & 1 & 1 &   &    \\
\rowcolor[cmyk]{0.05,0,0,0} &L\oplus E\oplus V         & 2 &   & 1 & 2 & 1 & 1 & 2 & 1 & 1  \\
\rowcolor[cmyk]{0.05,0,0,0} &L\oplus N\oplus V         & 2 &   &   & 2 &   & 2 & 1 & 1 & 1  \\
\rowcolor[cmyk]{0.05,0,0,0} &L\oplus N\oplus V\oplus E & 3 & 1 & 1 & 3 & 1 & 2 & 2 & 1 & 1
\end{array}$$
\caption{\em Techni-pion content of color neutral golden-class composite DM models.\label{TCpiDNA}}
\end{table}%

\subsection{Techni-pions at colliders}
As explained in eq.\eq{MDMs}, cosmology suggests two possible values for $M_{\rm DM}$:
$100 \TeV$ or $3\TeV$ depending on whether DM has a TCb asymmetry.
In both cases TCb DM is out of reach from LHC.  Furthermore,  DM production at  colliders gives missing energy signals which, 
especially at hadron colliders, can be undetectably below the neutrino background.

Composite DM models predict a richer collider phenomenology:
a general prediction is the existence of many resonances of various spin charged under the SM and with quantum numbers that can 
be determined from the ones of the constituents. Techni-pions are the lightest states in the theory so they are the most promising particles 
to be produced at colliders\footnote{Heavier spin-1 resonances  can  be singly produced through the mixing with  SM gauge bosons. 
They will then mostly decay in pairs of TC$\pi$.}.  
The anomalous coupling of some TC$\pi$ with SM vectors $V$  gives rise to single production of TC$\pi$, $V^* \to \pi V$ and $V^*V^*\to \pi$.
Techni-pions can also be produced in pairs via their SM gauge interactions,
with cross sections determined by their gauge quantum numbers and summarised e.g.\ in~\cite{MM}.
SM gauge interactions  and techni-quark masses  determine TC$\pi$ masses 
as in eq.~(\ref{mpigauge}) and (\ref{mpimass}). For an electro-weak triplet $3_0$ the gauge contribution alone 
is  $M_{3_0} \approx 0.1 M_{\rm DM}$ in a QCD-like $\SU(3)_{\rm TC}$. 

The two values for $M_{\rm DM}$,
$100 \TeV$ or $3\TeV$,  correspond to $M_{3_0}\approx 10\TeV$ (significantly above LHC capabilities)
or $M_{3_0} \approx 300\GeV$ (observable at LHC).

The only exception to the rule above is TC$\pi$ SM singlets $\eta$ that do not receive mass from SM gauge interactions. 
Their mass is entirely determined by the constituent techni-quark masses, such that  these TC$\pi$ could be very light. 
Usually such singlets undergo decays into pairs of SM gauge bosons through chiral anomalies 
\cite{sundrum}; when present their axion-like couplings to photons provides a mild constraint on their mass
(that need to be larger than a keV) and a production mechanism at colliders.

Each composite DM model predicts a distinctive set of TC$\pi$, as summarised in table~\ref{TCpiDNA}.

\medskip

The collider TC$\pi$ phenomenology can in principle discriminate golden-class from silver-class models \cite{ACDMcollider}.
In both cases TC$\pi$  without species number undergo anomalous decays into pairs of SM weak vectors, \beq  \pi_{1_0}, \pi_{3_0}, \pi_{5_0} \to WW, ZZ, \gamma \gamma \eeq
(models with coloured TC$\pi$, omitted from table~\ref{TCpiDNA}, also predict anomalous decays into gluon pairs).
In models with $G$-parity ($\Q=V$) the $\pi_{3_0}$ is stable.
Techni-pions made of different species decay via couplings that violate species number.

In silver-class models such couplings are provided by higher dimension operators involving SM particles
(for example 4-fermion operators),
giving decays into such SM particles. If these operators are suppressed by a large scale,  the decay is slow leading to 
displaced vertices or apparently stable particles on collider length scales, see \cite{sundrum} for a detailed discussion.

In golden-class models, species number and $G$-parity  can be broken by Yukawa couplings with the SM Higgs boson.
As a consequence, TC$\pi$ made of different species undergo decays into same specie TC$\pi$ (possibly off-shell) 
emitting one or more Higgs doublets $H$.  For example a doublet with  $Y=1/2$ and a singlet with $Y=1$ can decay as
\beq \pi_{2_{1/2}}\to H \pi_{1_0},\qquad \pi_{1_1} \to HH \pi_{1_0}
\eeq
and $\pi_{1_0}$ in turn decays into SM bosons through anomalies. 
Thereby, unlike in silver-class models, the SM fermions exhibit peaks in their invariant-mass distributions at the $h,W,Z$
masses (the Goldstone components of the Higgs doublet become the longitudinal components of the $W,Z$ vectors) \cite{ACDMcollider}.

\medskip

In models with Yukawa couplings the lighter technipions could also give interesting corrections to precision observables.
The loop corrections to electro-weak precision tests are universal and can be encoded in the $\hat S,\hat T,W,Y$ parameters~\cite{STWY}, 
that can be computed generalising section 2.1 of~\cite{MarS}, finding corrections of order $\alpha m_W^2/4\pi m_{TC\pi}^2$.
Concerning precision Higgs physics, $h\to \gamma\gamma$ gets corrected as~\cite{Djouadi}
\be
\frac{\Gamma(h \to \gamma\gamma)}{\Gamma(h \to \gamma\gamma)_{\rm SM}}
\simeq 1- 0.072
                \sum_{i}            \frac{v q_i^2   g_{hii}    }{m_{TC\pi_i}^2}
         \label{gamrat}
\ee
where $v\approx 174\GeV$ is the Higgs vev and $g_{hii}$ is the trilinear coupling between the Higgs boson
and the techni-pion mass eigenstate $i$ with electric charge $q_i$ and mass $m_{TC\pi_i}$ 
(here assumed to be much heavier than $m_h$, for simplicity).  The elemenatary Yukawa couplings $y \, H  \Q \Q $ give rise 
to cubic interactions such us $y \Lambda_{\rm TC}\, H \pi_2 \pi_{3}$ which generates $g_{hii} \sim y^2 \Lambda_{\rm TC}$ in the formula above. 
This effect might only be relevant in the asymmetric scenario where TC$\pi$ can be as light as few hundred GeV. 
A comprehensive study of precision observables will appear in \cite{ACDMcollider}.

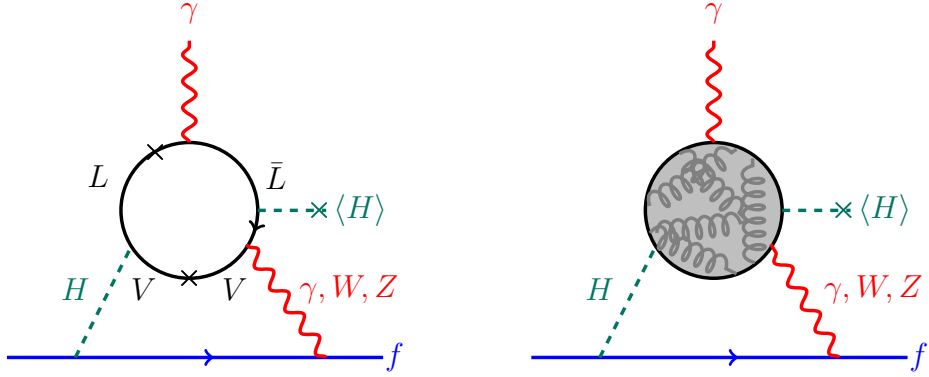
\begin{figure}[t]
\begin{center}
\begin{tikzpicture}[line width=1.3 pt, scale=3]
    {\color{blue}	\draw[fermion] (0,0)--(1.65,0);
    \node at (1.7,0) {$f$};}
	{\color{verdes}\draw[dashed] (0.3,0)--(0.55,0.5);
	\node at (0.3,0.3) {$H$};}
	\draw[fermion] (0.5,0.65) arc (180:-180:0.3);
	\node at (0.65,0.91) {$\times$};	\node at (0.65,0.91) {$\times$};\node at (0.65,0.91) {$\times$};
    \node at (0.4,0.8) {$L$};
	{\color{red}\draw[vector] (1.4,0)--(1.05,0.5);
	\node at (1.5,0.3) {$\gamma,W,Z$};}
{\color{red}	\draw[vector] (0.8,0.95)--(0.8,1.4);
	\node at (0.8,1.52) {$\gamma$};}
	\node at (0.8,0.35) {$\times$};	\node at (0.8,0.35) {$\times$};\node at (0.8,0.35) {$\times$};
	 \node at (1.18,0.8) {$\bar L$};
	{\color{verdes}\draw[dashed] (1.1,0.65)--(1.4,0.65);
	\node at (1.37,0.65) {$\times$};	\node at (1.37,0.65) {$\times$};\node at (1.37,0.65) {$\times$};}
	 \node at (0.6,0.3) {$V$};
	  \node at (1,0.3) {$V$};
{\color{verdes}	\node at (1.55,0.65) {$\langle H \rangle$};	}
 {\color{blue}	\draw[fermion] (2.3,0)--(3.95,0);
  \node at (4,0) {$f$};}
{\color{verdes}	\draw[dashed] (2.6,0)--(2.85,0.5);
	\node at (2.6,0.3) {$H$};}
	\draw[fill=lightgray] (2.8,0.65) arc (180:-180:0.3);
	\draw[gray][gluon] (2.82,0.67)--(3.2,0.92);
	\draw[gray][gluon] (3.28,0.42)--(3.27,0.88);
	\draw[gray][gluon] (2.82,0.55)--(3.2,0.37);
	\draw[gray][gluon] (2.95,0.9)--(3.25,0.6);
	\draw[gray][gluon] (2.85,0.55)--(3.25,0.6);
{\color{red}	\draw[vector] (3.65,0)--(3.35,0.5);
	\node at (3.8,0.3) {$\gamma,W,Z$};}
{\color{red}	\draw[vector] (3.1,0.95)--(3.1,1.4);
	\node at (3.1,1.52) {$\gamma$};}
	{\color{verdes}\draw[dashed] (3.4,0.65)--(3.7,0.65);
	\node at (3.67,0.65) {$\times$};	\node at (3.67,0.65) {$\times$};\node at (3.67,0.65) {$\times$};}
{\color{verdes}	\node at (3.85,0.65) {$\langle H \rangle$};	}
\end{tikzpicture}
\end{center}
\caption{\em Electron dipole moment generated by complex Yukawa couplings of techni-quarks, 
neglecting techni-color interactions (left) and including techni-color interactions (right).
\label{fig:edm2loops}}
\end{figure}

\subsection{Electron electric dipole}
Many models  contain  Yukawa couplings of techni-quarks with un-eliminable complex phases, generating
electric dipole moments for light SM fermions. 
Let us consider for example the model $\Q=L+V$ with $\SO(N)_{\rm TC}$. The techni-quark Lagrangian contains schematically\footnote{
The structure is  analogous to the Higgsino/wino system in split-supersymmetry~\cite{splitsusy2}. The same would work for $\SU(N)$ models with the difference that the triplet would be a Dirac fermion.},
\begin{equation}
m_L \bar L L+ \frac {m_V}2  V V + y_L  H^\dagger  V L+ y_R^* H   V  \bar L +\hbox{h.c.}
\end{equation}
It contains one physical CP-violating phase corresponding to ${\rm arg}\,[m_L m_V y_L^* y_R^*]$.
Ignoring technicolor interactions, an EDM is generated through the diagrams in the left panel of fig.~\ref{fig:edm2loops}, giving, in leading log approximation~\cite{splitsusy2},
\begin{equation}
d_f\sim  N e\,Q_f \frac {\alpha\, {\rm Im}[y_L^* y_R^*]} {16 \pi^3} \frac  {m_f}{m_L\, m_V} \ln\frac {m_L m_V}{m_H^2} \,.
\end{equation}
For the electron one finds
\begin{equation}
 {d_e}\sim  10^{-27}\, e\, {\rm cm}\times    {\rm Im}[y_L y_R] \times \frac{N}{ 3} \times\frac {\rm TeV^2}{m_L m_V}
\end{equation}
to be compared with the experimental bound $d_e<8.7 \times 10^{-29}~{\rm e\,cm}$ at 90\% C.L \cite{ACME}. 

However, the approximation of neglecting technicolor interactions is only reliable for $m_{L,V}>\Lambda_{\rm TC}$. 
In the more interesting regime $m_{L,V}<\Lambda_{\rm TC}$ techni-color  effects 
cannot be neglected and the loops will be dominated by the hadrons of the theory, as depicted in the right-handed panel of fig.~\ref{fig:edm2loops}.
A detailed study will appear in  \cite{ACDMcollider}.

\subsection{Gravitational waves}
Confining gauge theories can give rise to first order phase transitions. For $\SU(N)$ with $\NF$ massless flavours  this is believed
to happen in the window $3\le \NF \le 4N$ and $N>3$~\cite{Pis}. The phase transition occurs, within our framework, at a temperature 
$T\sim \Lambda_{\rm TC}$ (in the thermal dark matter scenario $ \Lambda_{\rm TC}\sim \mathcal{O}(10 {\rm \ TeV})$) and can lead to large anisotropic  fluctuations in the energy
momentum tensor sourcing the gravitational waves (GW) in the early universe. Following~\cite{Schwaller:2015tja}, we estimate the frequency of the peak in the GW signal as a function of the phase transition temperature $T$ as:
\beq
f_{\rm peak} = 3.3 \times 10^{-3}\ {\rm Hz} \times  \Big(\frac{T}{10\, {\rm TeV}}\Big) \times \Big( \frac{\beta}{10H}\Big)  
\eeq
where $\beta$ is the duration of the phase transition which is usually taken in the range 1-100 of a Hubble time $H$. 
For the reference values of the parameters, the amplitude of the expected GW signal is $h^2 \Omega_{\rm GW}\sim 10^{-9}$ ~\cite{Schwaller:2015tja} 
which is in the range that can be probed by future satellite experiments such as (E)LISA \cite{ELISA}. 

\begin{figure}[t]
$$
\includegraphics[width=0.45\textwidth]{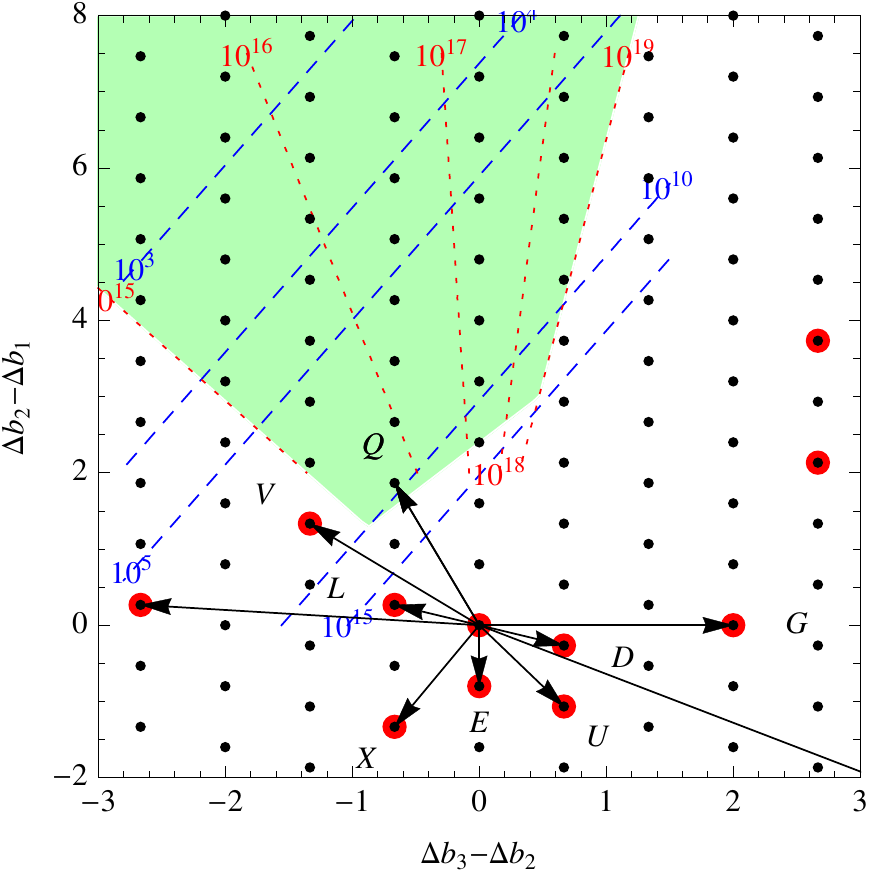}\qquad
\includegraphics[width=0.45\textwidth]{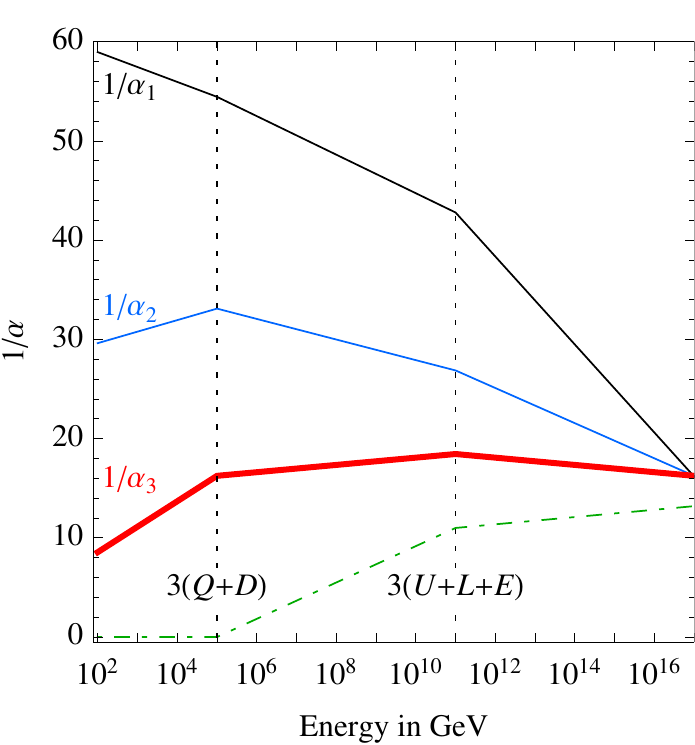}$$
\caption{\em {\bf Left}:  General analysis of unification, as described in the main text.
We show contour-lines of $M_{\rm GUT}/\GeV$ ($M_X/\Lambda_{\rm TC}$) in red (blue).
{\bf Right}:
Running of the gauge couplings  in the golden-class $\SU(3)_{\rm TC}$
model $\Q=Q\oplus \tilde{D}$. \label{fig:GUT}}
\end{figure}

\subsection{Unification of SM gauge couplings}
Throughout the paper we assumed that techni-quarks belong to fragments of unified $\SU(5)$ representations.
We here study if they can improve unification of SM gauge couplings.
The large number of independent masses allows for considerable freedom; we make the extra assumption
that the missing members of the unified $\SU(5)$ multiplets have a common mass $M_X$, below the GUT scale and above the TC scale $\Lambda_{\rm TC}$. Furthermore we make the rough assumption that the strong dynamics does not contribute to the running of the SM gauge couplings below the
$\Lambda_{\rm TC}\sim 100$ TeV, ignoring threshold effects including those of TC$\pi$.  With this mass ordering, in 1-loop approximation the running of gauge couplings is given by
\begin{equation}
\frac 1 {\alpha_i(M_Z)}=\frac 1 {\alpha_{\rm GUT}}+ \frac{b_i^{\rm SM}}{2\pi}\log \frac {M_{\rm GUT}}{M_Z} + \frac{\Delta b_i}{2\pi}\log \frac {M_X}{\Lambda_{\rm TC}}+ \frac{\Delta b}{2\pi}\log \frac {M_{\rm GUT}}{M_X} 
\end{equation}
where $\Delta b_i$ is the contribution from techni-quarks (listed in table~\ref{tab:unificaxion}), and $\Delta b$ is the contribution from the 
full SU(5) multiplets above $M_X$.
The three unification conditions determine the values of the three high-scale parameters $\alpha_{\rm GUT}$, $M_{\rm GUT}$ and $M_X$.
Inserting the SM values we find
\beq \ln\frac{M_X}{\Lambda_{\rm TC}} = \frac{68}{\Delta b_{21} - 1.9 \Delta b_{32}},\qquad
\ln\frac{M_{\rm GUT}}{M_X} = \frac{35.3 \Delta b_{21} - 49.2\Delta b_{32}}{\Delta b_{21} - 1.9 \Delta b_{32}}.
\eeq
Fig.~\ref{fig:GUT}a shows contour-values of $M_{\rm GUT}/\GeV$ (dotted red lines) and of $M_X/\Lambda_{\rm TC}$ (blue dashed lines)
as function of $\Delta b_{21} = \Delta b_2-\Delta b_1$ and of $\Delta b_{32} = \Delta b_3 - \Delta b_2$.
The dots in the figure are the grid of $\beta$-function coefficients allowed by SU(5) group theory~\cite{unificaxion},
and the arrows are the contributions to $(\Delta b_{32}, \Delta b_{21})$ from the fragments of $\SU(5)$ representations
listed in table~\ref{tab:unificaxion}.
The total $\beta$-function coefficient in any given model is obtained summing the contributions of each techni-quark 
taking into account their techni-color multiplicity $N$.

We see that models that can provide successful unification
must contain a $V$ or a $Q$ in order to obtain the desired sign of $\Delta b_{21}$.
For example:
\begin{itemize}

\item The golden-class $\SU(3)_{\rm TC}$ model $\Q=Q\oplus \tilde{D}$,
with techni-quarks coming from unified $5\oplus 10+\hbox{h.c.}$ multiplets of $\SU(5)$, provides successful unification
\begin{equation}
\alpha_{\rm GUT}\approx 0.06,\qquad ~M_{\rm GUT}\approx 2\times 10^{17} \ {\rm GeV},\qquad M_X\approx 2 \times 10^{11}\  {\rm GeV}
\times\frac{\Lambda_{\rm TC}}{100\TeV}
\end{equation}
having assumed {$\Lambda_{\rm TC}\approx100\TeV$}.
The running of the couplings is shown in fig.~\ref{fig:GUT}b.

\item 
The golden-class $\SO(3)_{\rm TC}$ model $\Q=V$, with $V$ coming from an adjoint of $\SU(5)$, provides
\begin{equation}
\alpha_{\rm GUT}\approx 0.065,\qquad ~M_{\rm GUT}\approx 3 \times 10^{14} \ {\rm GeV},\qquad M_X\approx 4 \times 10^{7}\  {\rm GeV}
\times\frac{\Lambda_{\rm TC}}{100\TeV}.
\end{equation}
Such a low unification scale would be excluded by proton decay. However, given the large uncertainties (we performed a one-loop analysis,
ignoring threshold effects that could be sizeable at the technicolor scale, in view of the light TC$\pi$)
such a model could still be viable. 

\item The silver-class $\SU(3)_{\rm TC}$ model with
$\Q=Q\oplus D\oplus U\oplus L$ coming from $\bar 5\oplus 10+\hbox{h.c.}$ multiplets of $\SU(5)$. We have,
\begin{equation}
\alpha_{\rm GUT}\approx 0.085,\qquad ~M_{\rm GUT}\approx M_X \approx 4\times 10^{17} \ {\rm GeV}
\end{equation}
having assumed $\Lambda_{\rm TC}\approx 200\TeV$.
The $DDU$ TCb can provide the observed Dark Matter, as discussed in appendix~\ref{sec:silvermodels},
model $\Q=D\oplus U$.

\end{itemize}

\section{Conclusions}\label{end}
Extensions of the SM with new strong interactions are interesting from the point of view of Dark Matter.
First, they naturally provide new stable particles, thanks to accidental symmetries analogous to baryon number
that guarantees the stability of the proton within the SM: DM could be the lightest techni-baryon (TCb) or techni-pion (TC$\pi$).
Second, the lightest among the many TCb tends to be the one with least SM gauge interactions, thereby
explaining why DM has no color, no electric charge, and at most a small hypercharge.

The models that we propose are compatible with all present bound from collider and precision experiments
because, with techni-quarks in a real representation of the SM gauge group, the new strong interactions do not break the electroweak symmetry.
The Higgs doublet is elementary and we do not address the hierarchy problem here. We use the old name `techni-color' in order to emphasize that 
we do not postulate desired good properties of effective Lagrangians.
On the contrary, we propose fundamental theories where
all the good properties follow from an appropriate choice of the quantum numbers: 
a concrete `techni-color'  gauge group and a concrete set of techni-quarks.

In the simplest `golden-class' of models, everything follows from a renormalizable Lagrangian.
In `silver-class' models, mild assumptions on non-renormalizable interactions are needed 
in order to break accidental symmetries and get rid of unwanted stable particles.
The list of `golden-class' models is meant to be exhaustive, within some assumptions:
no techni-scalars, only techni-fermions that transform in the fundamental representations of the technicolor gauge group, and
in representations of the SM gauge group which are compatible with SU(5) unification.
We found successful models with both $\SU(N)_{\rm TC}$ and $\SO(N)_{\rm TC}$ techni-color groups. We did not explore exceptional groups.

In $\SO(N)_{\rm TC}$ theories DM is a TCb, stable thanks to a $\Z_2 = {\rm O}(N)/\SO(N)$ symmetry: there is no conserved
techni-baryon number, such that DM is a real particle (a Majorana fermion for odd $N$, a real scalar for even $N$) with no TCb asymmetry,
no magnetic nor electric dipole. Assuming that its cosmological abundance comes from thermal freeze-out 
of techni-strong annihilations into TC$\pi$, the DM mass is expected to be around 100 TeV.
TCb mix once the Higgs boson acquires its vacuum expectation value (somehow analogously to the Wino/Bino/Higgsino system in supersymmetry), giving the following  phenomenology: in some regions of the parameter space DM can have an axial coupling to the $Z$, detectable in direct-detection signals;
in other regions of the parameter space it behaves as inelastic DM.

In $\SU(N)_{\rm TC}$ theories, the lightest TCb is a complex particle, stable thanks to conservation of
an accidental $\U(1)_{\rm TC}$ techni-baryon number. The DM mass could again be around 100 TeV: 
a Dirac fermion however can give sizable magnetic and electric dipole moments, giving direct-detection cross-sections 
enhanced in a characteristic way  at low recoil energy with respect to the case of a standard spin-independent cross section.
A large $\theta_{\rm TC}$-angle of the new strong sector can give an electric dipole such that direct detection is just below present bounds;
while a magnetic dipole cross section (suppressed at low DM velocities) is within the capabilities of future direct detection experiments.
Alternatively, the cosmological DM abundance could be due to a TCb asymmetry, with a DM mass around 3 TeV. 

In both cases, successful DM models often need Yukawa couplings with the Higgs boson
in order to break unwanted techni-flavor symmetries,
leading to extra spin-independent direct detection signals.   
CP-violating phases also lead to a possibly detectable electric dipole moment for the SM particles, such as the electron.

In some models composite DM has spin 1 or higher.

Concerning collider experiments, each model predicts a distinctive set of techni-pions, summarised in table~\ref{TCpiDNA}, which are
at most a factor 10 lighter than DM itself, than techni-baryons and than other vector composite resonances.
Some techni-pions undergo anomalous decays into SM vectors
(and can be singly produced via the inverse process), 
others decay into lighter techni-pions (and can be doubly produced via their SM gauge interactions)
emitting one or more Higgs doublets (i.e.\ $h,W,Z$), or, in silver-class models, emitting other SM particles.

\small

\subsubsection*{Acknowledgments}
This work was supported by the MIUR-FIRB grant RBFR12H1MW and by  the ESF grant MTT8.
We are grateful to Diego Becciolini for help with LieArt.
We thank  Tom DeGrand, M. D'Elia, E. del Nobile, E. Meggiolaro, A. Patella, Claudio Pica, R. Rattazzi, S. Rychkov, Yigal Shamir, for discussions.

\appendix

\section{Computing techni-baryons}\label{appendixA}
In section \ref{sec:SUN-pions-baryons} 
we computed  the multiplets of lighter TCb in $\SU(N)_{\rm TC}$ models.
The SM gauge interactions break explicitly the techni-flavor symmetry: here we outline how we compute the 
decomposition of the lightest TCb multiplet under the SM gauge group. 
We  label the SM quantum numbers of each  state under $\SU(3)_c \otimes \SU(2)_L \otimes \U(1)_Y$ as $(n_c, n_L)_Y$.

Let us first consider $\SU(3)_{\rm TC}$ theories with two species: $\Q=\Q_1+\Q_2$.  Models with more species can be solved by iteration. 
The lightest TCb fill a ${\tiny\Yvcentermath1  \yng(2,1)}$ representation of the techni-flavor group $\SU(d_1+d_2)_{\rm TF}$, where $d_{1,2}$ are the dimensions of the $\Q_{1,2}$ SM representations.
We  proceed in steps: first decompose the TCb multiplet under $\SU(d_1)\times \SU(d_2)$, with the embedding $ (d_1,1) \oplus (1,d_2)$, then decompose each component under the SM group and finally identify the $\SU(3)_c$ and $\SU(2)_L$ factors. 

From the first step we get:
\beq
{\tiny \Yvcentermath1 \yng(2,1) }=\big({\tiny\Yvcentermath1 \yng(1,1)\ ,\yng(1)} \big)_{2Y_1+Y_2}\oplus  \big({\tiny\Yvcentermath1  \yng(1) \ , \yng(1,1)}\big)_{2Y_2+Y_1} \oplus \big({\tiny\Yvcentermath1 \yng(2)\ ,\yng(1)}\big)_{2Y_1+Y_2} \oplus \big({\tiny\Yvcentermath1 \yng(1) \ , \yng(2)}\big)_{2Y_2+Y_1} \oplus \big({\tiny\Yvcentermath1  \yng(2,1) \ ,  1}\big)_{3Y_1} \oplus \big({\tiny\Yvcentermath1  1 \ ,\yng(2,1)}  \big)_{3Y_2} \,.
\label{general}
\eeq
The last two terms $\big({\tiny\Yvcentermath1  \yng(2,1) \ ,  1}\big)$, $\big({\tiny\Yvcentermath1  1 \ ,\yng(2,1)}  \big)$ correspond to TCb made only by $\Q_1$ or $\Q_2$ respectively and they reduce to one specie problems.  
The first four terms describe TCb composed  of both species. For example, $\big({\tiny\Yvcentermath1 \yng(1,1)\ ,\yng(1)} \big)_{2Y_1+Y_2}$ describes TCb made of $\Q_1 \Q_1 \Q_2$. 

\smallskip

We next decompose each component on the right hand side of eq.~\eqref{general} under the SM gauge group.
This can require non-trivial group theory computations: 
for example a techni-quark $V$ (triplet under $\SU(2)_L$) lies in the fundamental representation of techni-flavor $\SU(3)_{\rm TF}$:
TCb lie in higher representations of $\SU(3)_{\rm TF}$ that need to be decomposed under $\SU(2)_L$.
In general, we need to decompose a given representation with $K$ boxes of $\SU(n_c \, n_L)$  under \hbox{$\SU(n_c) \times \SU(n_L)$}, where the fundamental of $\SU(n_c \, n_L)$ is now embedded as $(n_c,n_L)$.  
This can be done writing all the representations of $\SU(n_c)$ and $\SU(n_L)$ with $K$ boxes. From group theory we know that each tableau is associated with a representation of the permutation group $S_K$ with a given symmetry.
Then $(D_1,D_2)$  appears in the decomposition if the product of $D_1$ and $D_2$ representations contains a component with the $S_K$ symmetry of the initial representation. 
Here is the decomposition of the two-index symmetric and antisymmetric tensors  under $\SU(n_c)$ and $\SU(n_L)$:
\beq
{\tiny \Yvcentermath1 \yng(1,1)} =\big({\tiny\Yvcentermath1 \yng(1,1)\ ,\yng(2)} \big)\oplus  \big({\tiny\Yvcentermath1  \yng(2) \ , \yng(1,1)}\big)  \qquad \qquad {\tiny \Yvcentermath1 \yng(2)} =\big({\tiny\Yvcentermath1 \yng(1,1)\ ,\yng(1,1)} \big)\oplus  \big({\tiny\Yvcentermath1  \yng(2) \ , \yng(2)}\big) \,.
\eeq
To be concrete, consider the techni-quark $\Q=(3,2)$. The decomposition of the two index tensors above under $\SU(3)_c \times \SU(2)_L$ become:
\beq
{\tiny \Yvcentermath1 \yng(1,1)} =(6,1)\oplus (\bar{3},3),  \qquad \qquad {\tiny \Yvcentermath1 \yng(2)} =(6,3)\oplus (\bar{3},1).
\eeq
If with respect to any SM group factor the techni-quarks transform in a representation $n_i$ higher than the fundamental, we can embed it into the fundamental of $\SU(n_i)$ and decompose representations of this larger group under the SM group.
For example in the $\Q=V$ model, the techni-quark is a vector of $\SU(2)_L$: we can think of the two-index symmetric $3$ of $\SU(2)_L$ as the fundamental of $\SU(3)$ into which $\SU(2)$ is embedded symmetrically.
With simple group algebra we find:
\begin{eqnarray}
\SU(3) &:& 3 \times 3 = 6 \oplus \overline{3} \,, \quad 3 \times \overline{3} = 1 \oplus 8 \notag \\
\SU(2) &:&  3 \times 3 = 1 \oplus 3 \oplus 5 \,,
\end{eqnarray}
from which we get the decomposition rules $6= 5 \oplus 1$ and $8 = 5 \oplus 3$ for the $\SU(3)$ representations under the $\SU(2)$ group.


After this step, each state in eq.~\eqref{general} is labeled by the quantum numbers $(n_{c1}, n_{L1}, n_{c2}, n_{L2})_{Y}$. To obtain the final representation under the SM group we have to identify $\SU(3)_c$ and $\SU(2)_L$ factors, taking the tensor product $n_{c1} \otimes n_{c2}$ and $n_{L1} \otimes n_{L2}$.

\bigskip

For $\SU(4)_{\rm TC}$,
we can proceed analogously. First, we decompose the lightest TCb multiplet of $\SU(d_1+d_2)$ under  $\SU(d_1)\times \SU(d_2)$:
\beq
{\tiny \Yvcentermath1 \yng(2,2) }=\big({\tiny\Yvcentermath1 \yng(2,1)\ ,\yng(1)} \big)_{3Y_1+Y_2} \oplus  \big({\tiny\Yvcentermath1  \yng(1,1) \ , \yng(1,1)}\big)_{2Y_2+2Y_1} \oplus \big({\tiny\Yvcentermath1 \yng(2)\ ,\yng(2)}\big)_{2Y_1+2Y_2} \oplus \big({\tiny\Yvcentermath1 \yng(1) \ , \yng(2,1)}\big)_{Y_1+3Y_2} \oplus \big({\tiny\Yvcentermath1  \yng(2,2) \ ,  1}\big)_{4 Y_1} \oplus \big({\tiny\Yvcentermath1  1 \ ,\yng(2,2)}  \big)_{4Y_2} \,,
\eeq
then we decompose each representation under the SM group and identify the $\SU(3)_c$ and $\SU(2)_L$ factors respectively.

\medskip

As discussed in section \ref{sec:SON-pions-baryons}, the $\SO(N)_{\rm TC}$ theories can be analyzed starting from the results of the $\SU(N)_{\rm TC}$ models.

\section{Silver-class composite DM models}\label{sec:silvermodels}

We  here list silver-class $\SU(N)_{\rm TC}$ and $\SO(N)_{\rm TC}$ models restricted for simplicity to $N=3,4$ techni-colors and $N_s \le 2$ species of techni-quarks. These models  satisfy TC asymptotic freedom and do not give rise to sub-Planckian Landau poles. But, besides to acceptable DM candidates, they give rise to unwanted stable states, that are TC$\pi$ with hypercharge or color, stable because of accidental symmetries such as species number or $G$-parity.
They can be made unstable with extra model building, for example adding higher dimension operators that break the accidental symmetries, as explained in section \ref{sec:breaking}. 

\subsection{SU$(N)_{\rm TC}$ silver-class models}\label{modelLE}

\medskip

\subsubsection*{SU($N$)$_{\rm TC}$ model $\Q = N \oplus E$}
This model has $N_S=\NF=2$. TC$\pi$ fill the adjoint of $\SU(2)_{\rm TF}$:
\beq
\label{NE-TCpi}
\hbox{TC$\pi$ : } 3 = 1_{0,\pm 1} \qquad \hbox{under $\SU(2)_L\otimes\U(1)_Y$}.
\eeq
TC$\pi$ made by both techni-quarks have non zero hypercharge and are stable because of species number. If we want to make the model phenomenologically viable, we need to break species number by ad hoc assumptions.
For $N=3$, the lightest TCb live in the fundamental of $\SU(2)_{\rm TF}$, that is
\beq
\hbox{TCb : }  2 = 1_{\pm 1} \qquad \hbox{under $\SU(2)_L \times \U(1)_Y$}\,.
\eeq
The DM candidate is the spin $3/2$ singlet $NNN^*$ that belongs to the symmetric representation ${\tiny \yng(3)}$ of $\SU(2)_{\rm TF}$. It can be the lightest TCb if $m_N \ll m_E$.
The same conclusion is valid for $N=4$, where 
the DM candidate is the spin $2$ singlet $NNNN^*$ that lives in the symmetric representation ${\tiny \yng(4)}$ of $\SU(2)_{\rm TF}$.

\subsubsection*{SU($N$)$_{\rm TC}$ model $\Q = E \oplus \tilde E$}
This model with $\NF=2$ can give rise to a neutral TCb for $N=4$. 
It presents a Landau pole for $g_Y$ slightly above the Planck scale and gives rise to the following TC$\pi$
\beq
\hbox{TC$\pi$ : } 3 = 1_{0,\pm 2} \quad \hbox{under $\SU(2)_L \otimes \U(1)_Y$} \,.
\eeq
The $1_{\pm 2}$ TC$\pi$ made by both species are stable, so that we need to break species number.
The model gives only one lighter TCb, that is a SM singlet made by $EE\tilde{E}\tilde{E}$ and is a good DM candidate.

\subsubsection*{SU($N$)$_{\rm TC}$ model $\Q = L \oplus \tilde L$}
The TC$\pi$ of this model with $\NF=4$ are:
\beq
\hbox{TC$\pi$ : } 15 = 1_{0,\pm 1} \oplus 3_{2 \times 0,\pm 1}  \quad \hbox{under $\SU(2)_L \otimes \U(1)_Y$}
\eeq
where states with hypercharge are stable, unless the species number symmetry is broken.
Analogously to the previous model, it can provide a DM candidate for $N=4$, where the lighter TCb fill a $20'$ of $\SU(4)_F$, that decomposes as
\beq
\hbox{TCb : } 20' = 1_{2 \times 0,\pm 1,\pm 2} \oplus 3_{0,\pm 1} \oplus 5_0 \quad \hbox{under $\SU(2)_L \otimes \U(1)_Y$} \,.
\eeq
The list contains two singlets $LL\tilde{L}\tilde{L}$ that are good DM candidates.

\subsubsection*{SU($N$)$_{\rm TC}$ model $\Q = L \oplus E$}
This model, studied in detail in section~\ref{QCDlike}, has $\NF=3$ and for $N=3$ gives rise to the successful DM candidate $LLE$. 
In this case, both TC$\pi$ and TCb live in the adjoint of $\SU(3)_{\rm TF}$, that decomposes as
\beq
8 = 1_0 \oplus 2_{\pm 3/2} \oplus 3_0 \quad \hbox{under $\SU(2)_L \times \U(1)_Y$} \,.
\eeq
TC$\pi$ made of $L\bar E$ are $2_{\pm 3/2}$ states, stable because of the unbroken species number symmetry. 
One can get rid of the unwanted stable particles by ad-hoc model building, breaking accidental symmetries with higher dimensional operators or adding new particles. For example one can add a scalar doublet $H'$ with $|Y|={3/2}$ such that the Yukawa coupling $H'L  E$ is allowed.

The TCb DM candidate is the singlet $LLE$.
As explained in section \ref{QCDlike}, techni-quark masses favor $LLL$ or $LEE$ as the lightest state, 
so that the $LLE$ singlet can be the stable DM candidate if gauge interactions contribute to mass splitting more than techni-quark masses.

\subsubsection*{SU($N$)$_{\rm TC}$ model $\Q = V \oplus E$}
This model with $\NF=4$ is allowed only for $N=3$ and gives rise to the following TC$\pi$:
\beq
\hbox{TC$\pi$ : } 15 = 1_0 \oplus 3_{0,\pm 1} \oplus 5_0 \quad \hbox{of $\SU(2)_L \otimes \U(1)_Y$} 
\eeq
with the $3_{\pm 1}$ states stable because of species number. 
The lightest multiplet of TCb decomposes as 
\beq
\hbox{TCb : } \overline{20} = 1_{1} \oplus 3_{0,1,2} \oplus 5_{0,1} \quad \hbox{under $\SU(2)_L \otimes \U(1)_Y$}
\eeq
and the DM candidate is the triplet made by $VVV$.

\subsubsection*{SU($N$)$_{\rm TC}$ model $\Q = D \oplus N$, $\Q = U \oplus N$ and $\Q = Q \oplus N$}
We can study together the first two models ($\NF=4$) defining $Y=1/3,-2/3$ for $D$ and $U$ respectively. 
For the $\Q = D(U) \oplus N$ models we get the following TC$\pi$:
\begin{eqnarray}
\label{DN,UN-TCpi}
\hbox{TC$\pi$ : } 15 &=& 1_0 \oplus \overline{3}_Y \oplus 3_{-Y} \oplus 8_0  
\quad \hbox{of $\SU(3)_c \otimes \U(1)_Y$}\,.
\end{eqnarray}
Analogously, for the $\Q= Q \oplus N$ ($\NF=7$) model we get:
\begin{eqnarray}
\label{QN-TCpi}
\hbox{TC$\pi$ : } 48 &=& (1,1)_0 \oplus (1,3)_0 \oplus (3,2)_{1/6} \oplus (\overline{3},2)_{-1/6} \oplus (8,1)_0 \oplus (8,3)_0   
\end{eqnarray}
of $\SU(3)_c \otimes \SU(2)_L \otimes \U(1)_Y$.
Because of species number symmetry, we have stable TC$\pi$ with color and/or hypercharge, so that we need to break this accidental symmetry to avoid the strong experimental bounds.
As in the other models containing the singlet $N$, the TCb DM candidate is an higher spin state made only by $N$ techni-quarks. 
For $N=3$ and $N=4$ it has spin $3/2$ and $2$ respectively and it can be the lightest if the other techni-quark is sufficiently heavier.
 
\subsubsection*{SU($N$)$_{\rm TC}$ model $\Q = D \oplus V$, $\Q = U \oplus V$}
As before, we study together the two models defining $Y=1/3,-2/3$ for $D$ and $U$ respectively.
Because of the presence of the techni-quark $V$, these models are allowed only for $N=3$. They have $\NF=6$ and give rise to the following states:
\begin{eqnarray}
\hbox{TC$\pi$ : } 35 &=& (1,1)_0 \oplus (1,3)_0 \oplus (\overline{3},3)_Y \oplus (3,3)_{-Y} \oplus (1,5)_0 \oplus (8,1)_0  \nonumber \\
\hbox{TCb : } 70 &=& (\overline{3},1)_Y \oplus (1,3)_0 \oplus (\overline{3},3)_Y \oplus (3,3)_{2Y} \oplus (1,5)_0 \oplus (\overline{3},5)_Y \oplus (8,1)_{3Y} \oplus (\overline{6},3)_{2Y}  
\end{eqnarray}
under $\SU(3)_c \otimes \SU(2)_L \otimes \U(1)_Y$.
We need to break species number in order to avoid stable TC$\pi$ made by both species that are colored and have hypercharge. The DM candidate is the triplet $VVV$: from the DM point of view these models are a trivial extension of the $\Q = V$ model described in section \ref{sec:SU3-models}.

\subsubsection*{SU($N$)$_{\rm TC}$ model $\Q = D \oplus U$}
This model has $\NF=6$, so the TC$\pi$ live in the adjoint of $\SU(6)_{\rm TF}$:
\beq
\hbox{TC$\pi$ : } 35 = 1_{0,\pm 1} \oplus 8_{2 \times 0, \pm 1} \quad \hbox{under $\SU(3)_c \otimes \U(1)_Y$}\,.
\eeq
TC$\pi$ with non zero species number are colored and have hypercharge, so we need to break the species number symmetry.
The model gives a good TCb DM candidate for $N=3$, where the lightest TCb are
\beq
\hbox{TCb : } 70 = 1_{0,-1} \oplus 8_{2 \times 0,1, 2 \times (-1),-2} \oplus \overline{10}_{\,0,-1} \quad \hbox{under $\SU(3)_c \otimes \U(1)_Y$}\,,
\eeq
and the DM candidate is the singlet made by $D D U$.

\subsection{SO$(N)_{\rm TC}$ silver-class models}\label{}

\medskip


\subsubsection*{SO($N$)$_{\rm TC}$ model $\Q=E$}
This model has $\NF=2$ and it is free from Landau poles up to $N=8$.
The unwanted stable TC$\pi$ are singlets with $Y=\pm2$.
The model can provide a TCb DM candidate for even $N$.
For $N=4$, there is only one TCb with spin 0 that is a singlet and thus a good DM candidate.


\subsubsection*{SO($N$)$_{\rm TC}$ model $\Q=E \oplus N$}
This $\NF=3$ model does not allow for Yukawa couplings and contains unwanted stable TC$\pi$ with hypercharge, that belongs to the $5$ of $\SO(3)_{\rm TF}$:
\beq
\hbox{TC$\pi$ : } 5 = 1_{0,\pm 1,\pm 2} \quad \hbox{of  $\SU(2)_L\otimes\U(1)_Y$} \,.
\eeq
The model can be extended up to $N=8$. 
For $N=3$ the lightest TCb is a singlet $NE \bar E$ that lives in the ${\tiny \yng(1)}$ representation of $\SO(3)_{\rm TF}$, while the heavier TCb are
\beq
\hbox{TCb : } {\tiny \yng(2,1)} = 5 = 1_{0,\pm 1,\pm 2} \quad \hbox{of  $\SU(2)_L\otimes\U(1)_Y$} \,.
\eeq
For $N=4$, the DM candidate is a singlet linear combination of $E \bar E N N$ and $E \bar E E \bar E$. In this case the ${\tiny \yng(2,2)}$ representation is absent, so the full set of TCb is already specified.

\subsubsection*{SO($N$)$_{\rm TC}$ model $\Q=E \oplus V$}
In this model $\NF=5$ so that TC$\pi$ compose a $14$ of $\SO(5)_{\rm TF}$ that is
\beq
\hbox{TC$\pi$ : } 14 = 1_{0,\pm 2} \oplus 3_{\pm 1} \oplus 5_0 \quad \hbox{of  $\SU(2)_L\otimes\U(1)_Y$} \,,
\eeq
where the states with hypercharge, made by $E E$, $\bar E \bar E$ or $E V$, $\bar E V$ are stable. Extra assumptions are needed to break the accidental symmetries and remove unwanted stable states. The model is valid up to $N =7$. For $N=3$ the TCb DM candidate is the $3_0$ state $V(E \bar E + VV)$ belonging to the representation ${\tiny \yng(1)}$ of the unbroken flavor group.
The other TCb are:
\beq
\hbox{TCb : } {\tiny \yng(2,1)}= 35 = 1_{\pm 1} \oplus 3_{2 \times 0,\pm 1, \pm 2} \oplus 5_{0,\pm 1} \quad \hbox{of  $\SU(2)_L\otimes\U(1)_Y$} \,.
\eeq
For $N=4$ the DM candidate is a singlet $(E \bar E + VV)^2$, while the remaining TCb are given by 
\beq
\hbox{TCb : } {\tiny \yng(2,2)} = 35 = 1_0 \oplus 3_{0,\pm 1} \oplus 5_{0,\pm 1,\pm 2} \quad \hbox{of  $\SU(2)_L\otimes\U(1)_Y$} \,,
\eeq
plus a set of states living in the same representations as the TC$\pi$ above.

\subsubsection*{SO($N$)$_{\rm TC}$ model $\Q=L$}
This model has $\NF=4$,
the TC$\pi$ lie in a $9$ of $\SO(4)_{\rm TF}$ that decomposes as
\beq  
\hbox{TC$\pi$ : } 9=3_{\pm1}\oplus 3_0 \qquad\hbox{under $\SU(2)_L\otimes\U(1)_Y$}. 
\eeq
TC$\pi$ made by $LL$ and $\bar L \bar L$ have non zero hypercharge and are stable because of accidental $\U(1)$ species symmetry.
The extra physics needed to avoid unwanted stable TC$\pi$ can be nicely realised 
considering the golden-class model $\Psi=L\oplus N$ in the limit where $m_N \gg \Lambda_{\rm TC}$, such that the
$(LH)^2$ effective operator is generated at low energy.

TCb can contain a DM candidate for $N$ even.
For $N=4$, this is the singlet $(L \bar L)^2$. The other TCb that need to be specified are
\beq 
\hbox{TCb : } 
{\tiny \yng(2,2)} = 10=5_0\oplus 1_{\pm 2,\pm 1,0} \qquad\hbox{under $\SU(2)_L\otimes\U(1)_Y$}
\, .
\eeq
Landau poles are avoided for $N\le 14$.


\subsubsection*{SO($N$)$_{\rm TC}$ model $\Q=L\oplus E$}
This model with $\NF=6$ allows for two Yukawa couplings, leaving an unbroken U(1)
species number, rotating $L,\bar E$ with a common phases, and $\bar L, E$ with the opposite phase.
Thereby TC$\pi$ made by $L  \bar E$, $ \bar L E$ are stable and have hypercharge $\pm 3/2$. The full list of TC$\pi$ is:
\beq 
\hbox{TC$\pi$ : } 20= 1_{\pm 2,0}\oplus
2_{\pm 3/2, \pm 1/2}\oplus 3_{\pm 1, 0} \qquad\hbox{under $\SU(2)_L\otimes\U(1)_Y$}\,. 
\eeq 
Again, we need to break the unwanted accidental symmetry in some way. 
The techni-color theory is asymptotically free only for $N\ge 4$ and sub-Planckian Landau poles are avoided for $N\le 5$.
The model gives a singlet TCb DM candidate for $N=4$, that is $(L \bar L + E \bar E)^2$. To obtain the full list of TCb we need to decompose the multiplet of heavier TCb
under $\SU(2)_L\otimes\U(1)_Y$:
\beq 
\hbox{TCb : } {\tiny \yng(2,2)} = 84=5_0 \oplus 4_{\pm 3/2, \pm 1/2}
\oplus 3_{\pm 3, \pm 2, 2\times (\pm 1), 2\times 0}
\oplus 2_{\pm 5/2, 2\times(\pm 3/2), 3\times (\pm 1/2)}
\oplus 1_{\pm 2, 2 \times (\pm 1), 3\times 0} .
\eeq

%

\subsubsection*{SO($N$)$_{\rm TC}$ models $\Q=U$, $\Q=D$ }
We study the two models jointly, defining  $Y=-2/3$ for $\Q=U$ and  $Y=1/3$ for $\Q=D$.
The model is asymptotically free for $N \ge 4$ and Landau poles are avoided  up to $N=6$  for $U$ and to $N=14$ for $D$.
In both cases $\NF=6$ and TC$\pi$ lie in the $20$ of $\SO(6)_{\rm TF}$ that decomposes as
\beq  
\hbox{TC$\pi$ : } 20=8_0 \oplus \Bigl( \overline{6}_{2Y} \oplus {\rm h.c.} \Bigr) \qquad \hbox{under $\SU(3)_c\otimes\U(1)_Y$}. 
\eeq
Because of the $\U(1)$ accidental symmetry there are unwanted stable colored TC$\pi$ made by $D \bar D$ or $U \bar U$.
The models provide singlets TCb DM candidates only for even $N$.
For $N=4$ the full TCb list contains the multiplet 
\beq 
\hbox{TCb : } 
{\tiny \yng(2,2)} = 84=1_0 \oplus 8_{0} \oplus 27_0 \oplus \Bigl(3_{2Y} \oplus 6_{4Y} 
 \oplus 15_{2Y} \Bigr)  \quad\hbox{under $\SU(3)_c\otimes\U(1)_Y$}.
\eeq

\subsubsection*{SO($N$)$_{\rm TC}$ models  $\Q=D \oplus N$, $\Q=U \oplus N$}
A trivial extension of the previous models is given by the $N \oplus D$ and $N \oplus U$ models, with $\NF=7$. They give rise to an extended list of TC$\pi$:
\beq  
\hbox{TC$\pi$ : } 27= 1_0 \oplus  8_0  \oplus \Bigl( 3_{-Y} \oplus 6_{-2Y} \oplus {\rm h.c} \Bigr) \qquad \hbox{under $\SU(3)_c\otimes\U(1)_Y$} \,, 
\eeq
where the extra state made by $N$ and $D$ or $U$ are unwanted stable particles.
For $N=4$ the lightest TCb DM candidate is again a singlet, made by $D \bar D(D \bar D + NN)$ or $U \bar U(U \bar U + NN)$.

\subsubsection*{SO($N$)$_{\rm TC}$ models $\Q=D \oplus V$, $\Q=U \oplus V$}
These are less trivial extensions of the $D$ and $U$ models, with $\NF=9$. The list of TC$\pi$ is
\beq
\hbox{TC$\pi$ : } 44 = (1,1)_0 \oplus (1,5)_0 \oplus (8,1)_0 \oplus \Bigl( (3,3)_{-Y} \oplus  (6,1)_{-2 Y} \oplus \,{\rm h.c.} \Bigr) \eeq
under $\SU(3)_c \otimes \SU(2)_L \otimes \U(1)_Y$,
including stable unwanted states with color and hypercharge.
Asymptotic freedom requires $N \ge 4$, the model can be extended up to $N=7$ for $D \oplus V$ and up to $N=6$ for $U \oplus V$.
For $N=4$, the lightest TCb DM candidate is a singlet $(D \bar D+ VV)^2$ or $(U \bar U+VV)^2$, while the heavier TCb contain the multiplet
\begin{eqnarray}
\hbox{TCb : } {\tiny \yng(2,2)} &=& 495 = 2 \times (1,1)_0 \oplus (1,3)_0 \oplus (1,5)_0 \oplus 2 \times (8,1)_0 \oplus  (8,3)_{0,3Y} \oplus (8,5)_0 \oplus (27,1)_0 \nonumber \\
&& \oplus 
\Bigl( (3,1)_{2Y} \oplus (3,3)_{2Y,2 \times -Y} \oplus (\overline{6},1)_{2Y,-4Y} \oplus (3,5)_{-Y} \oplus (\overline{6},3)_{-Y} \oplus  (\overline{6},5)_{2Y} \nonumber \\
&& \oplus  (15,1)_{2Y} \oplus (15,3)_{-Y} \oplus  {\rm h.c.} \Bigr) \quad \hbox{under $\SU(3)_c \otimes \SU(2)_L \otimes \U(1)_Y$} \,.
\end{eqnarray}

\subsubsection*{SO($N$)$_{\rm TC}$ models $\Q=D \oplus E$}
This model with $\NF=8$ is valid from $N=4$ up to $N=6$, while the analogous model $U \oplus E$ suffers by a sub-Planckian Landau pole for $g_Y$.
The decomposition of the TC$\pi$ multiplet under $\SU(3)_c \otimes \U(1)_Y$ is 
\beq
\hbox{TC$\pi$ : } 35 = 1_{2 \times 0,\pm 2} \oplus 8_0 \oplus \Bigl( 3_{2/3,-4/3} \oplus 6_{-2/3} \oplus  {\rm h.c.} \Bigr) \,.
\eeq
The list includes stable TC$\pi$ with color and hypercharge, so that we need to break the accidental symmetries to remove unwanted stable particles.
For $N=4$ the DM candidate is a TCb singlet $(D \bar D + E \bar E)^2$. To complete the list of TCb we need the decomposition of the multiplet
\begin{eqnarray}
\hbox{TCb : } {\tiny \yng(2,2)} &=& 300 = 3 \times 1_0 \oplus 8_{5\times 0, 2 \times (\pm 2)} \oplus 27_0 \oplus \Bigl( 3_{4 \times 2/3, 2 \times (-4/3)} \ \nonumber \\
&&\oplus \overline{6}_{8/3,2 \times 2/3, 3 \times (-4/3)}  \oplus 15_{2 \times 2/3,-4/3} \oplus {\rm h.c.}\Bigr)
\end{eqnarray}
under $\SU(3)_c \otimes \U(1)_Y$.

\subsubsection*{SO($N$)$_{\rm TC}$ models $\Q=D \oplus L$ and $\Q=U \oplus L$}
These models can be analysed jointly defining $Y=1/3$ and $Y=-2/3$ for $D$ and $U$ respectively.
They are characterized by $\NF=10$, the model with $D$ is allowed for $4 \le N \le 9$, while the model with $U$ is allowed only for $N=4$. 
TC$\pi$ fill a $54$ dimensional representation of $\SO(10)_{\rm TF}$:
\beq
\hbox{TC$\pi$ : } 54 = (1,1)_0 \oplus (1,3)_{0,\pm 1} \oplus (8,1)_0  \oplus \Bigl( (3,2)_{1/2-Y,-1/2-Y} \oplus (\overline{6},1)_{2Y} \oplus \hbox{h.c.} \Bigr) 
\eeq 
under $\SU(3)_c \otimes \SU(2)_L \otimes \U(1)_Y$. Accidentally stable TC$\pi$ have color and/or hypercharge, so that we need to break these accidental symmetries to make the model phenomenologically viable. 
The lightest TCb for $N=4$ is a singlet $(D \bar D + L \bar L)^2$ or $(U \bar U+L \bar L)^2$ and it is a good DM candidate. The set of heavier TCb contains the following states 
\begin{eqnarray}
\hbox{TCb : } 770 &&= (1,1)_{4\times 0,2\times (\pm 1),\pm 2} \oplus (1,3)_{2 \times 0,\pm 1} \oplus (1,5)_0 \oplus (8,1)_{3 \times 0,\pm 1} \oplus (8,2)_{\pm (1/2+3Y),\pm (1/2-3Y)} \nonumber \\
&&\oplus (8,3)_{2 \times 0, \pm 1}  \oplus (27,1)_0 \oplus  \Bigl( (3,1)_{1+2Y,2 \times 2Y,-1+2Y} \oplus (3,2)_{3/2-Y,3 \times(1/2-Y),3 \times(-1/2-Y),-3/2-Y} \nonumber \\
&& \oplus (3,3)_{2Y} \oplus (\overline{6},1)_{2Y,-4Y} \oplus (3,4)_{1/2-Y,-1/2-Y} \oplus (\overline{6},2)_{1/2-Y,-1/2-Y} \oplus (\overline{6},3)_{1+2Y,2Y,-1+2Y} \nonumber \\
&&\oplus (15,1)_{2Y} \oplus (15,2)_{1/2-Y,-1/2-Y} \oplus {\rm h.c.} \Bigr) \,,
\end{eqnarray}
under $\SU(3)_c \otimes \SU(2)_L \otimes \U(1)_Y$.

\subsubsection*{SO($N$)$_{\rm TC}$ models $\Q=G \oplus N$ and $\Q=G \oplus E$}
These are simple extensions of the $\Q=G$ model described in section \ref{sec:SON-models}, with $\NF=9$ and $\NF=10$ respectively, allowed only for $N=4$. Because of species number TC$\pi$ made by different species are stable and since they have hypercharge and/or color, they are excluded by DM direct search bounds. We need ad hoc assumptions to break the accidental symmetry and make them unstable.
The lists of TC$\pi$ for the $G \oplus N$ and $G \oplus E$ models respectively are:
\begin{eqnarray}
\hbox{TC$\pi$ : } 44 &&= 1_0 \oplus 2 \times 8_0 \oplus 27_0 \\
\hbox{TC$\pi$ : } 54 &&= 1_{0,\pm 2} \oplus 8_{0,\pm 1} \oplus 27_0 \quad \hbox{under $\SU(3)_c \otimes \U(1)_Y$} \,.
\end{eqnarray}
In both cases, the DM candidate is a singlet, in the first model it is made by $GG(GG+NN)$, in the second by $(GG + E \bar E)^2$.
Here we present the decomposition under $\SU(3)_c \otimes \U(1)_Y$ of the heavier TCb multiplet for the $G \oplus N$ and $G \oplus E$ models respectively:
\begin{eqnarray}
\hbox{TCb : } {\tiny \yng(2,2)} &=& 495 = 1_0 \oplus 4\times 8_0 \oplus 6 \times 27_0 \oplus 64_0 \oplus \Bigl( 2 \times 10_0 \oplus 28_0 \oplus 2 \times 35_0 \oplus {\rm h.c} \Bigr) \\
\hbox{TCb : } {\tiny \yng(2,2)} &=& 770 = 2 \times 1_0 \oplus 8_{\pm 2,3\times (\pm 1),0} \oplus 27_{\pm 2, 2\times (\pm 1),4 \times 0} \oplus 64_0 \oplus \Bigl(10_{\pm 1, 2 \times 0} \oplus 28_0 \nonumber \\
&& \oplus 35_{\pm 1, 0} \oplus {\rm h.c.} \Bigr) \,.
\end{eqnarray}

\end{document}